\documentclass[aps,prb,twocolumn,showpacs,superscriptaddress]{revtex4}
\usepackage{amssymb}
\usepackage{amsfonts}
\usepackage{amsmath}
\usepackage{bm}
\usepackage{textcomp}
\usepackage{color}
\usepackage{longtable}
\usepackage{graphicx}
\usepackage{dcolumn}

\usepackage[a4paper=true,pagebackref=false]{hyperref}
\hypersetup{colorlinks,citecolor=blue,filecolor=blue,linkcolor=blue,urlcolor=blue}

\begin{document}

\title{Ferromagnetism in semiconductors and oxides: prospects from a ten years' perspective}

\author{Tomasz Dietl} \email{dietl@ifpan.edu.pl}
\affiliation{Institute of Physics, Polish Academy of Science,
  al.~Lotnik\'ow 32/46, PL-02-668 Warszawa, Poland}
\affiliation{Institute of Theoretical Physics, University of Warsaw,
  PL-00-681 Warszawa, Poland}

\date{\today}

\begin{abstract}
Over the last decade the search for compounds combining the resources of semiconductors and ferromagnets has evolved into an important field of materials science. This endeavour has been fuelled by continual demonstrations of remarkable low-temperature functionalities found for ferromagnetic structures of (Ga,Mn)As, p-(Cd,Mn)Te, and related compounds as well as by ample observations of ferromagnetic signatures at high temperatures in a number of non-metallic systems. In this paper, recent experimental and theoretical developments are reviewed emphasising that, from the one hand, they disentangle many controversies and puzzles accumulated over the last decade and, on the other, offer new research prospects.
\end{abstract}


\maketitle
\section*{Introduction}

Advances in the epitaxy of semiconductor compounds have made it possible to fabricate quantum structures in which confined electrons or photons exhibit outstanding properties and functionalities. Similarly, the atomic precision of metal and oxide film deposition has allowed to master a number of striking spin transport phenomena.  The discovery of
ferromagnetism in p-type Mn-doped IV-VI (ref.~\onlinecite{Story:1986_PRL}), III-V
(refs~\onlinecite{Ohno:1992_PRL,Ohno:1996_APL,Esch:1997_PRB}),  and II-VI (refs~\onlinecite{Haury:1997_PRL,Ferrand:2000_JCG}) compounds has opened a road towards the development of multifunctional materials systems bridging the resources of semiconductor quantum structures and ferromagnetic
multilayers as well as has enabled the study of collective magnetic phenomena as a function of the spin and carrier densities.

Over the last ten years or so the field of ferromagnetism in dilute magnetic semiconductors (DMSs) and dilute magnetic oxides (DMOs) has evolved into an important branch of materials science. The comprehensive research on these systems has been stimulated by continual demonstrations of outstanding low-temperature functionalities in (Ga,Mn)As, p-(Cd,Mn)Te,
and related structures\cite{Awschalom:2007_NP,Spintronics:2008_B}, some
examples being spin-injection\cite{Ohno:1999_N}, electric-field\cite{Ohno:2000_N,Chiba:2008_N} and electric-current\cite{Chernyshov:2009_NP} control of
magnetism, tunneling anisotropic magnetoresistance in planar junctions\cite{Gould:2004_PRL} and in the Coulomb blockade regime\cite{Wunderlich:2006_PRL}, as well as current-induced domain displacement without the assistance of a magnetic field\cite{Yamanouchi:2004_N}. These findings have put into focus the interplay of magnetization texture and dynamics with carriers' population
and currents, a broad topic of today's physics of spintronic materials. At the same time, since the first report on (Ti,Co)O$_2$ (ref.~\onlinecite{Matsumoto:2001_S}), the persistence of spontaneous magnetization to above the room temperature has been found for a number of DMOs and DMSs, and even for materials nominally containing no transition
metal (TM) impurities.

However, despite the massive investigations, the origin and control of ferromagnetism in DMSs and DMOs is, arguably, the most controversial research topic in today's materials science and the condensed matter physics. As emphasized here, the abundance of contradicting views has resulted from intertwined theoretical and experimental challenges, requiring the application of cutting edge computational and materials
nanocharacterisation methods, often becoming available only now. In this way, DMSs and DMOs emerge as an outstanding playground to test our understanding of unanticipated relationships between growth conditions and a self-organized alloy structure\cite{Bonanni:2010_CSR} as well as between
quantum localization, carrier correlation, and ferromagnetism, and constitute an active research
direction\cite{Sheu:2007_PRL,Sawicki:2010_NP,Richardella:2010_S}.

We begin this review by recalling the foundations of the $p-d$ Zener model, proposed a decade ago to describe the origin and properties of ferromagnetism in p-type Mn-doped semiconductors\cite{Dietl:2000_S}. A
crucial role of the Anderson-Mott localization in the physics of these systems is then discussed in the context of other models put forward to explain outstanding findings which have been accumulated over the recent years for (Ga,Mn)As.  It is shown that the $p-d$ Zener model describes a number of thermodynamic and micromagnetic properties of III-V and II-VI
DMSs as well as continues to constitute a good starting point to address the question about prospects of research on hole-mediated ferromagnetic semiconductors. The second part of the review is devoted to the origin and control of high temperature ferromagnetism in DMSs and DMOs. We argue,
exploiting results of extensive nanocharacterisation works, that puzzling properties of these compounds reflect a highly non-random distribution of magnetic cations. While a number of appealing functionalities has already been demonstrated for these ferrmagnetic/semiconductor nanocomposites, we
are only at the beginning of the road to demonstrate device structures of these emerging materials systems.

\section*{THE $p-d$ ZENER MODEL}

In view of the progress in materials fabrication by epitaxial methods\cite{Ohno:1992_PRL,Ohno:1996_APL,Esch:1997_PRB,Matsukura:1998_PRB,Haury:1997_PRL,Ferrand:2000_JCG} it was timely a decade ago to understand the ferromagnetism in DMSs as well as to ask whether the Curie temperature $T_{\mathrm{C}}$ can be raised to above 300~K from the 110~K observed at that time in (Ga,Mn)As containing only 5\% of Mn (ref.~\onlinecite{Matsukura:1998_PRB}).  Photoemission\cite{Okabayashi:1998_PRB} as well as optical studies in the single impurity limit\cite{Linnarsson:1997_PRB} demonstrated that Mn provides both localised spins and itinerant holes mutually coupled by a $p-d$ exchange interaction.  Zener\cite{Zener:1951_PR} first proposed the model of ferromagnetism driven by the exchange interaction between band carriers and localized spins.  In the case of semiconductors, the Zener model\cite{Dietl:1997_PRB,Jungwirth:1999_PRB} is equivalent to the approach developed by Ruderman, Kittel, Kasuya, and Yosida (RKKY), in which the Friedel oscillations of the spin density are taken into account\cite{Dietl:1997_PRB}.

In the proposed implementation of the Zener model the structure of the valence subbands was described by the Kohn-Luttinger six bands' $kp$ hamiltonian, taking the spin-orbit interaction into account\cite{Dietl:2000_S}. Thermodynamic characteristics were then evaluated in the mean-field approximation. At the same time, arguments were presented why the model is valid even on the insulator side of the Anderson-Mott metal-to-insulator transition (MIT), provided that the holes remain weakly localised.

This approach was found to constitute an appropriate minimal theory, capable to describe adequately the magnitude of $T_{\mathrm{C}}$ and of magnetic anisotropy fields induced by biaxial strains in (Ga,Mn)As and p-(Zn,Mn)Te (ref.~\onlinecite{Dietl:2000_S}).
It was also pointed out that GaN and ZnO containing appropriately high concentrations of {\em both} Mn spins ($x \gtrsim 5$\%) and delocalised or weakly localised holes in the valence band ($p \gtrsim 3.5\cdot 10^{20}$~cm$^{-3}$) might support the ferromagnetic order to above the room temperature. It was underlined, however, that prior to the verification of this prediction, important issues of solubility limits and self-compensation as well as of the transition to a strong-coupling case with the decreasing lattice constant need to be addressed experimentally\cite{Dietl:2000_S}.

\section*{A GUIDE THROUGH OTHER MODELS}

\begin{figure}
\includegraphics[width=9cm]{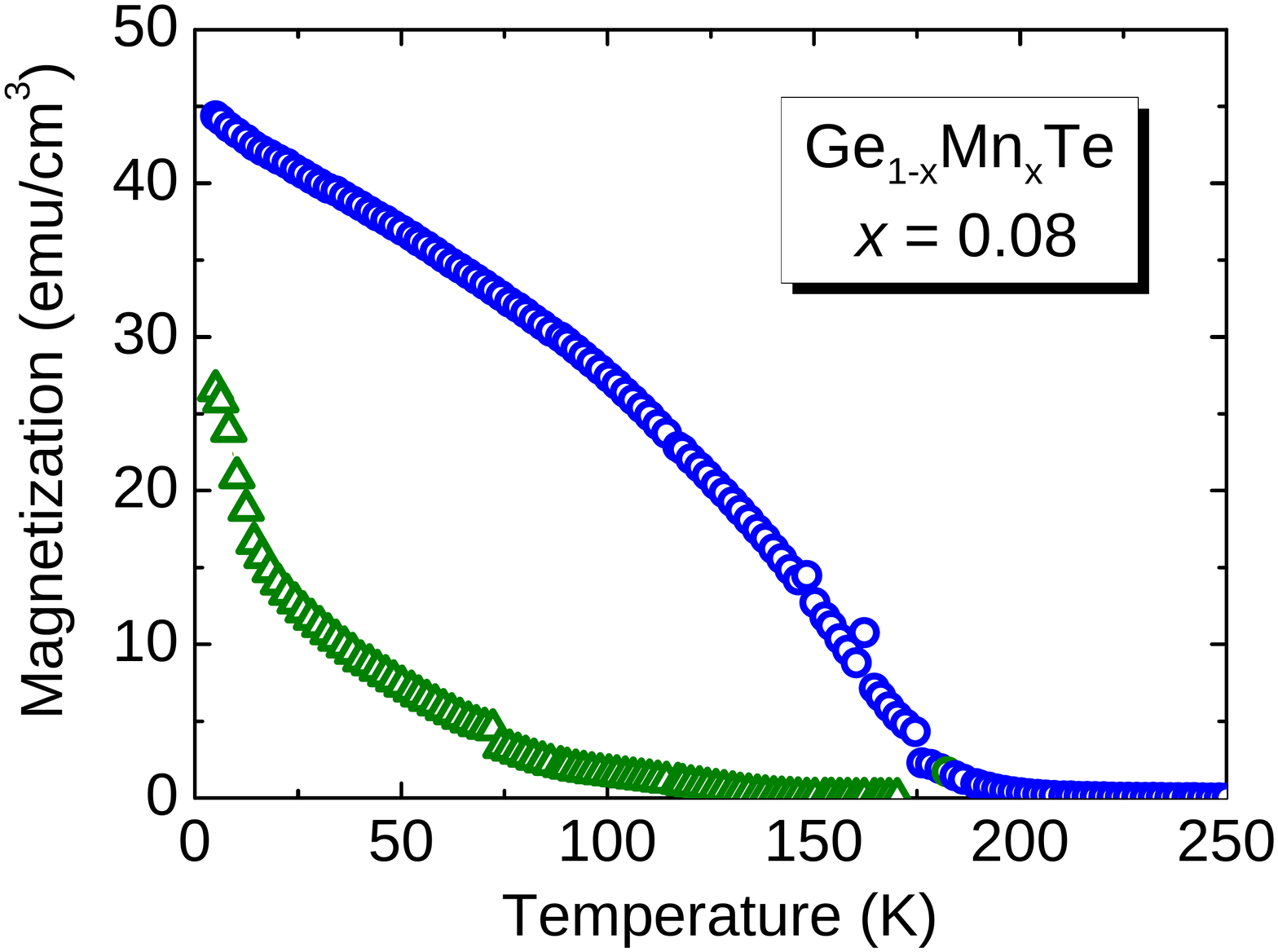}
\includegraphics[width=9cm]{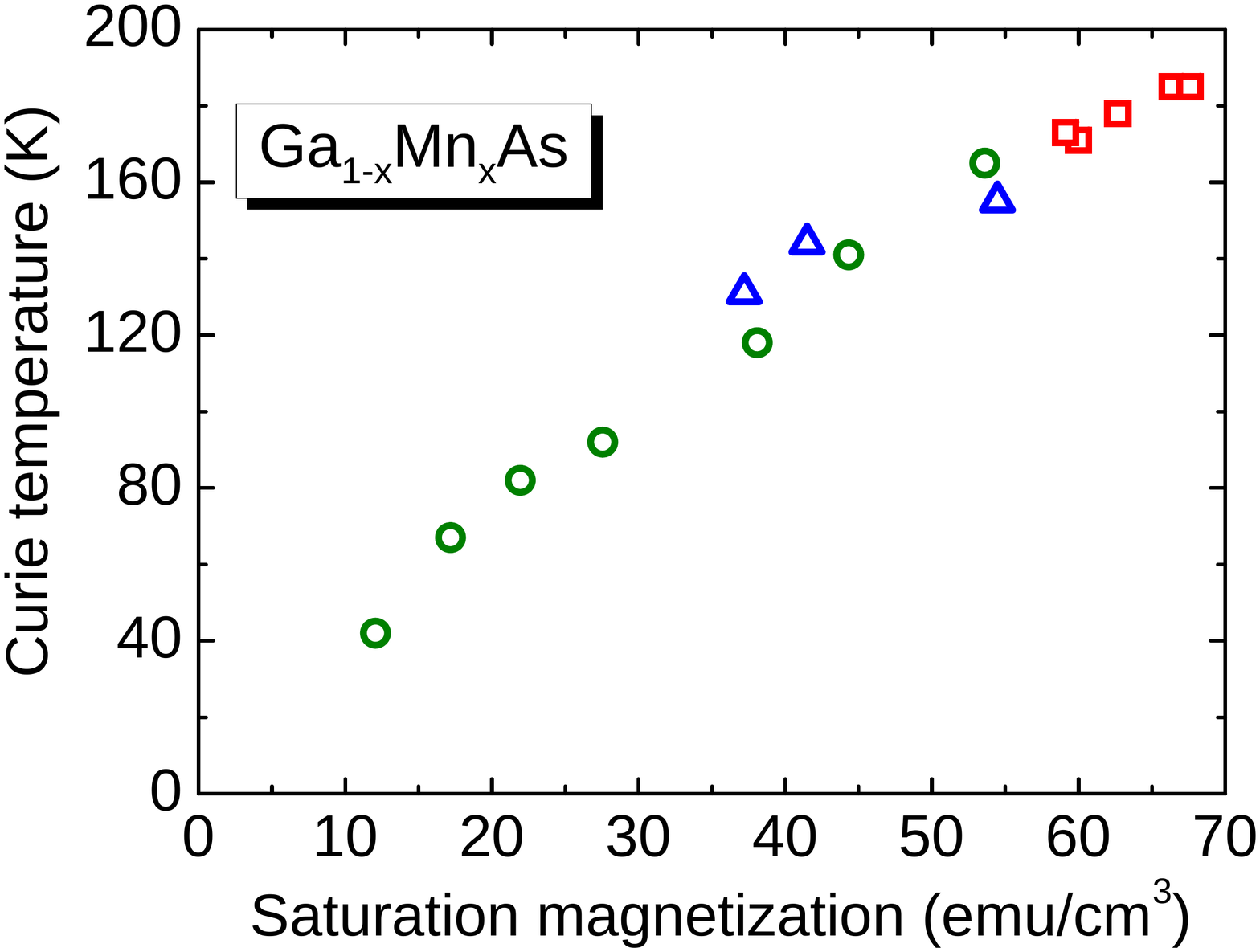}
\caption{Experimental data for p-type DMSs films showing the Curie temperature $T_{\mathrm{C}}$ approaching 200~K at the effective Mn concentration $x_{\mathrm{eff}}$ below 10\%. Upper panel: temperature dependence of magnetization in (Ge,Mn)Te with high (circles) and low (triangles) hole concentrations (after~ref.~\onlinecite{Fukuma:2008_APL}); Lower panel: $T_{\mathrm{C}}$ as a function of saturation magnetization $M_{\text{Sat}}$ for annealed (Ga,Mn)As films grown in various molecular beam epitaxy (MBE) systems (after~ref.~\onlinecite{Wang:2008_APL}).}
\label{fig:GeMnTe_GaMnAs}
\end{figure}

Extensive studies over last ten years have made clear that ferromagnetic DMSs and DMOs form two distinct classes. The first class comprises p-type Mn-based DMSs, in which the ferromagnetism is associated with the presence of holes. Here, step by step improvements in growth protocols and in post-grown processing have made it possible to increase the Mn and hole densities, particularly in (Ge,Mn)Te (ref.~\onlinecite{Fukuma:2008_APL}) and (Ga,Mn)As (ref.~\onlinecite{Olejnik:2008_PRB,Wang:2008_APL,Chen:2009_APL}), in which the magnitudes of $T_{\mathrm{C}}$'s approach now 190~K at a value of the effective Mn concentration $x_{\mathrm{eff}}$ below 10\%, as implied by the magnitude of saturation magnetization (see Fig.~\ref{fig:GeMnTe_GaMnAs}). While this evolution of $T_{\mathrm{C}}$ is consistent with the $p-d$ Zener model, its basic foundation, namely that in the concentration range relevant to ferromagnetism the holes reside in the valence band in (Ga,Mn)As and related systems has been objected by two schools of thoughts:

\begin{itemize}
\item Following pioneering {\em ab initio} work carried out for (In,Mn)As (ref.~\onlinecite{Akai:1998_PRL}), it has been argued based on the outcome of available first principles methods that the holes reside in band-gap states derived from the TM $d$ levels, so that the relevant spin-spin coupling mechanism is {\em the double-exchange}\cite{Sato:2003_EPL,Mahadevan:2004_APL}.

\item A series of findings from optical and transport studies, and hard to reconcile with expectations for the holes moving in a weakly perturbed valence band, have been taken as an evidence for the location of the Fermi energy within a Mn-acceptor {\em impurity band} detached from the valence band or retaining the $d$ character of Mn dopants even in the region, where they overlap with the valence band on the metallic side of the MIT \cite{Burch:2008_JMMM,Alberi:2008_PRB}.
\end{itemize}

To the second class of ferromagnetic systems belongs a broad range of semiconductors, oxides, and carbon derivatives, showing ferromagnetic-like features persisting to above room temperature without the presence of itinerant holes or, in some cases, even without intentional doping by TM impurities\cite{Liu:2005_JMSME,Coey:2006_COSSMS,Bonanni:2007_SST}. A number of diverging models has been proposed to explain the origin of this intriguingly robust ferromagnetism:

First, in a long series of {\em ab initio} works a ferromagnetic ground state has been found for dilute TM spins even in the absence of band carriers\cite{Sato:2010_RMP}. Following pioneering contributions\cite{Blinowski:1996_PRB,Akai:1998_PRL}, the effect has been assigned to the ferromagnetic superexchange\cite{Blinowski:1996_PRB} or, more frequently, to the double exchange\cite{Akai:1998_PRL} that turns up, within the employed computation methodologies, if the states derived from the TM $d$ levels are partly occupied for one spin direction\cite{Sato:2010_RMP}.

Second, it is supposed that electrons either residing in the conduction band\cite{Walsh:2008_PRL} or forming bound magnetic polarons\cite{Coey:2005_NM} mediate ferromagnetic couplings between dilute TM spins. Alternatively, the presence of these couplings is assigned to carriers residing on defects, such as vacancies\cite{Wang:2009_PRB} or on residual impurities, such as hydrogen\cite{Park:2005_PRL}.

Third, the ferromagnetic response is related to spins of electrons residing on point or extended defects and coupled by an exchange interaction. Within this so-called $d^0$ model of high temperature ferromagnetism, the presence of magnetic impurities is either unnecessary or serves merely to bring the Fermi level to the relevant defect states\cite{Coey:2008_JPD}.

Finally, it has been persistently suggested\cite{Dietl:2003_NM,Bonanni:2007_SST} that the limited solubility of TMs' impurities in particular hosts may result in the formation of nanoscale regions containing a large density of magnetic cations and, thus, specified by a high spin ordering temperature.

\section*{WHERE DO THE HOLES RESIDE IN {(Ga,Mn)As}?}

According to the double exchange scenario\cite{Sato:2003_EPL,Mahadevan:2004_APL}, the anti-crossing picture\cite{Alberi:2008_PRB,Cho:2008_APL,Mayer:2010_PRB}, and the impurity band models\cite{Esch:1997_PRB,Burch:2008_JMMM}, the hole states retain impurity band characteristics in the density regime relevant to the ferromagnetism of (Ga,Mn)As. For a comprehensive presentation of arguments in favour of such a scenario we remit the readers to ref.~\onlinecite{Burch:2008_JMMM}.

Another view, shared by the present author and exposed in detail elsewhere\cite{Jungwirth:2007_PRB,Dietl:2008_JPSJ}, is that similarly to other doped semiconductors  the carrier localization in p-type DMSs results from a collective effect of randomly distributed scattering centres upon the Fermi liquid of strongly correlated band carriers\cite{Altshuler:1985_B,Lee:1985_RMP,Belitz:1994_RMP,Dietl:2008_JPSJ}. Guided by previous extensive studies of non-magnetic semiconductors, we may anticipate that rather than absolute values, only certain scaling characteristics of d.c., a.c., and tunneling conductivity tensors can be presently interpreted theoretically near the MIT, at least at temperatures below the momentum relaxation rate. This is in contrast to thermodynamic properties, such as electronic specific heat, which are virtually unperturbed by disorder and electronic correlation at the localization boundary\cite{Altshuler:1985_B,Lee:1985_RMP,Belitz:1994_RMP}.

\begin{figure}[ht]
\begin{center}
\includegraphics[width=10cm]{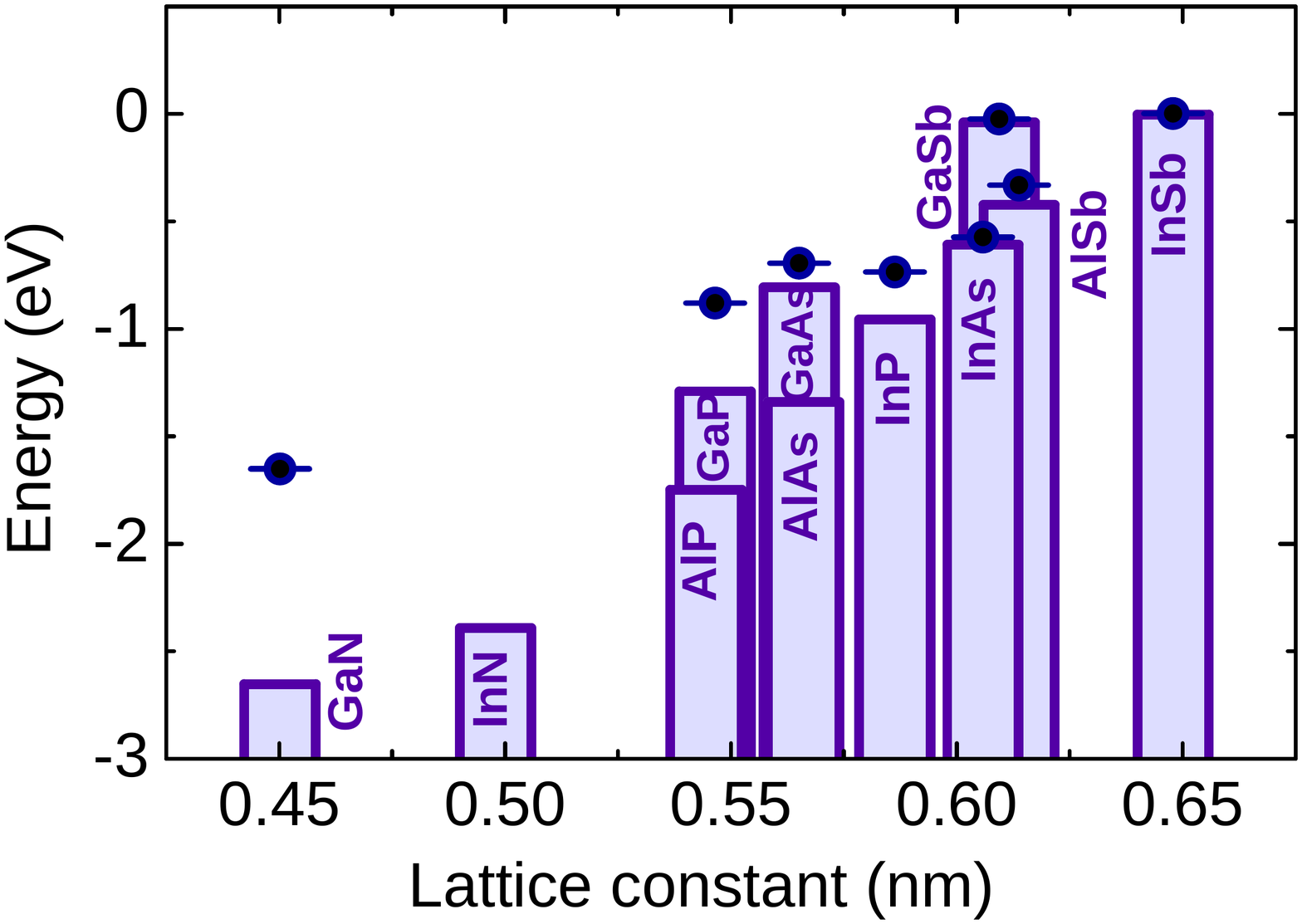}
\caption[]{Experimental energies of Mn levels in the gap of III-V compounds with respect to the valence-band edges (after ref.~\onlinecite{Dietl:2002_PRB}).}
 \label{fig:EI}
\end{center}
\end{figure}

Empirically, the Anderson-Mott MIT occurs for the carrier concentration $p_{\mathrm{c}}$ at which the magnitude of the kinetic energy per band carrier, $E_{\mathrm{kin}} \approx (3/5)E_{\mathrm{F}}$, calculated with no disorder diminishes to about one third of the single impurity binding energy $E_{\mathrm{I}}$ (ref.~\onlinecite{Edwards:1978_PRB}). In Fig.~\ref{fig:EI} the experimental values of $E_{\mathrm{I}}$ for Mn acceptors in various III-V semiconductors are shown\cite{Dietl:2002_PRB}. As seen, $E_{\mathrm{I}}$ and, thus, $p_{\mathrm{c}}$ is enhanced rather dramatically comparing to non-magnetic acceptors, particularly on going from antimonides to nitrides through arsenides and phosphides. This shift is caused by the $p-d$ hybridisation, whose importance grows with the decreasing cation-anion bond length, ultimately resulting in a transition to the strong coupling limit, where the hole binding is dominated by the $p-d$ interaction\cite{Dietl:2008_PRB}.

According to the scaling theory and relevant experiments\cite{Altshuler:1985_B,Lee:1985_RMP,Belitz:1994_RMP,Dietl:2008_JPSJ}, the Anderson-Mott MIT is continuous. Hence, the carrier localization length $\xi$ decreases rather gradually from infinity at the MIT towards the impurity Bohr radius in the strongly localised regime, so that at a length scale smaller that $\xi$, the wave function retains an extended character. Such band-like carriers, whose quantum diffusivity vanishes at the MIT, were originally thought to mediate the long-range interactions between the TM spins in DMSs in the whole density regime relevant to ferromagnetism\cite{Dietl:2000_S,Dietl:1997_PRB}.

In agreement with this view temperature dependent quantum corrections to the conductance\cite{Neumaier:2009_PRL} and the character of {\em tunneling} DOS\cite{Richardella:2010_S} are consistent in (Ga,Mn)As with the expectations for Anderson-Mott localization of holes in the GaAs valence band. In particular, scanning tunneling microscopy data\cite{Richardella:2010_S}, though affected presumably by the proximity to the surface, point rather directly to the crucial importance of {\em both} disorder and carrier correlation in the relevant range of Mn concentrations. At the same time, the direct visualisation\cite{Richardella:2010_S} of spatial fluctuations in local DOS provides a support to the view\cite{Sawicki:2010_NP} that the disappearance of ferromagnetism with carrier localization proceeds {\em via} an intermediate superparamagnetic-like phase.

Within the Anderson-Mott localization model, the effects of disorder and carrier correlation appear as some broadening and Landau's renormalisation of valence band {\em thermodynamic} DOS at $E_{\mathrm{F}}$, $\rho_{\mathrm{F}}$ which determines $T_{\mathrm{C}}$ within the $p-d$ Zener model (refs \onlinecite{Dietl:2000_S,Dietl:1997_PRB,Boukari:2002_PRL}). In this context,  thermoelectric power $S$ at high temperatures is relevant, as it is a good measure of $\rho_{\mathrm{F}}$, so that its magnitude may tell between the valence band and impurity band pictures.  Recent measurements of $S$ in compensated (Ga,Mn)As were interpreted in terms of the anti-crossing model treating broadening of the impurity band as an adjustable parameter\cite{Mayer:2010_PRB}. In this way,  a rather small difference between the magnitudes of $S$ for (Ga,Mn)As and GaAs:Be at given hole densities was explained.  We note, however, that the Mott formula $S= (\pi^2/3e)k_{\mathrm{B}}^2T \partial_{E_{\mathrm{F}}} \ln\sigma(E_{\mathrm{F}})$ with $\rho_{\mathrm{F}}$ of the GaAs valence band\cite{Neumaier:2009_PRL}  describes quantitatively not only the data\cite{Mayer:2010_PRB} for GaAs:Be but also for (Ga,Mn)As assuming that the energy dependence of the apparent hole mobility $\mu = \sigma/ep$ changes from the value expected for acoustic phonon scattering, $\mu \sim E_{\mathrm{F}}^{-1/2}$, to the one specific for ionised impurity scattering,  $\mu \sim E_{\mathrm{F}}^{3/2}$, when the degree of compensation increases.

Furthermore, within the impurity-band models, the magnitude of $T_{\mathrm{C}}$ is predicted to reach a maximum,  when the Fermi level is shifted across the peak in the density of the impurity states. The accumulated data for (Ga,Mn)As show that the value of $T_{\mathrm{C}}$ decreases monotonically when diminishing the hole concentration by gating\cite{Nishitani:2010_PRB} or by increasing the concentration of compensating donors\cite{Mayer:2010_PRB,MacDonald:2005_NM,Edmonds:2002_APL}. On the other hand, it was found in recent studies that  the magnitude of  $T_{\mathrm{C}}$ actually {\em increases} by co-doping with  Si donors\cite{Cho:2008_APL} and, moreover, it goes through a maximum as a result of the modulation doping by Be acceptors\cite{Cho:2008_SST}.  As underlined in these works, the findings are consistent with the impurity band scenario. However, the presented data reveal that the increase of $T_{\mathrm{C}}$ in Si-doped samples and its decrease in the Be case is associated with, respectively,  an enlargement and a reduction of the saturation value of magnetization and, thus, of the Mn concentration $x_{\mathrm{eff}}$ that determines the magnitude of $T_{\mathrm{C}}$. The results\cite{Cho:2008_APL,Cho:2008_SST} can, therefore, be explained by the known anticorrelation\cite{Furdyna:2004_JPCM} between $x_{\mathrm{eff}}$ and the density of holes during the epitaxy, here reduced by Si donors\cite{Cho:2008_APL} or "siphoned off" from the Be-doped barrier into the grown quantum well\cite{Cho:2008_SST}.

At the same time, low values of  $\mu$ (below 10~cm$^2$/Vs), taken as an evidence for the large magnitude of an effective hole mass\cite{Burch:2008_JMMM,Alberi:2008_PRB}, can be explained by the proximity to the MIT, where charge diffusion is much reduced by quantum localization effects. Furthermore, referring to a shift of an a.c. conductivity maximum with the hole density\cite{Burch:2008_JMMM}, which contradicts the expectations of the Drude-Boltzmann theory, we emphasise  that the frequency dependent conductance near the MIT, at least up to frequencies of the order of the momentum relaxation rate, is dominated by quantum localization effects\cite{Altshuler:1985_B,Lee:1985_RMP,Belitz:1994_RMP}, whose presence may account for the observed anomalies.

\section*{CURIE TEMPERATURE FOR CARRIER-MEDIATED FERROMAGNETISM IN III-V {DMSs}}

\begin{figure}[ht]
\includegraphics[trim = 2cm  0 0 0, clip, width=9cm]{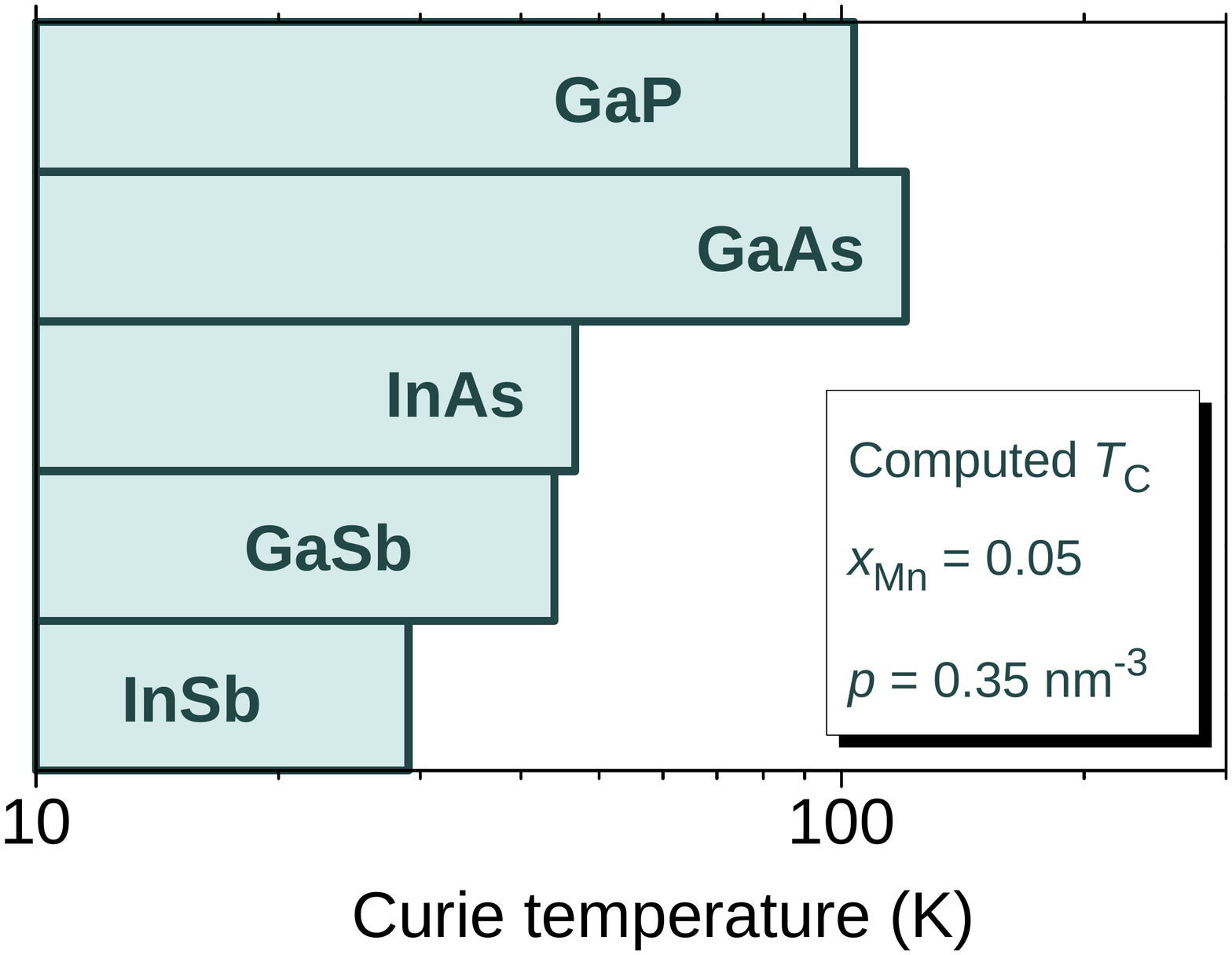}
\includegraphics[trim = 2cm  0 0 1cm, clip, width=9cm]{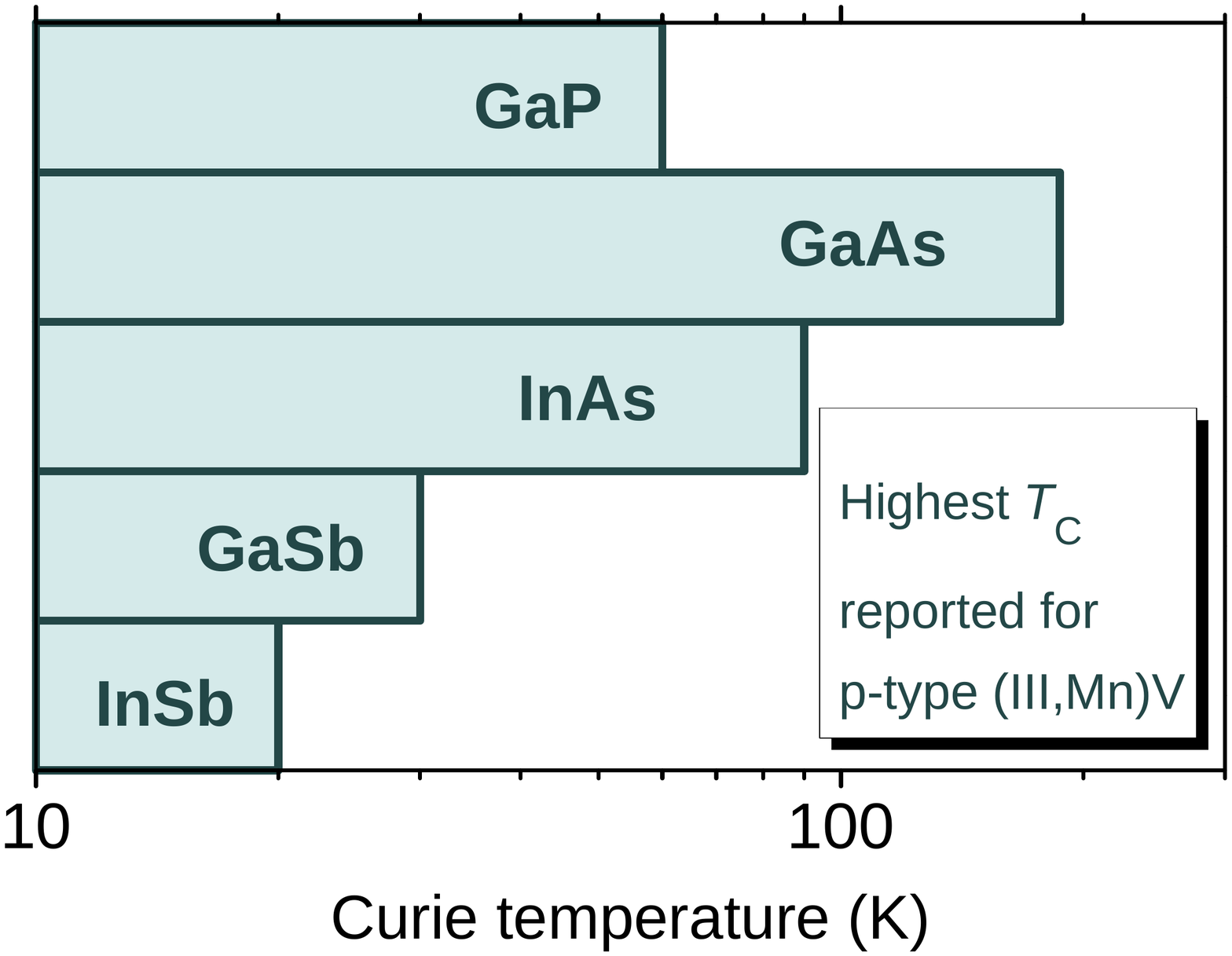}
\caption{Predictions of the $p-d$ Zener model compared to experimental data for p-type (III,Mn)V DMSs. Upper panel: computed values of the Curie temperature $T_{\mathrm{C}}$ for various p-type semiconductors containing 5\% of Mn and $3.5\times 10^{20}$ holes per cm$^3$ (after ref.~\onlinecite{Dietl:2000_S}; the value for (In,Mn)Sb is taken from ref.~\onlinecite{Jungwirth:2002_PRB}). Lower panel: the highest reported values for (Ga,Mn)P (ref.~\onlinecite{Scarpulla:2005_PRL}); (Ga,Mn)As (ref.~\onlinecite{Olejnik:2008_PRB,Wang:2008_APL}); (In,Mn)As (ref.~\onlinecite{Schallenberg:2006_APL}); (Ga,Mn)Sb (ref.~\onlinecite{Abe:2000_PE}); (In,Mn)Sb (ref.~\onlinecite{Wojtowicz:2003_APL}).}
\label{fig:TC}
\end{figure}

In Fig.~\ref{fig:TC} the highest values of $T_{\mathrm{C}}$ found to date in p-type Mn-based III-V DMSs are reported\cite{Scarpulla:2005_PRL,Olejnik:2008_PRB,Wang:2008_APL,Chen:2009_APL,Schallenberg:2006_APL,Abe:2000_PE,Wojtowicz:2003_APL}, and compared to the early predictions of the $p-d$ Zener model\cite{Dietl:2000_S,Jungwirth:2002_PRB} for fixed values of the Mn and hole concentrations. We see that the theory reproduces the chemical trends and describes semi-quantitatively the absolute values of $T_{\mathrm{C}}$. The observed trend reflects a decrease of the $p-d$ exchange energy for larger cation-anion distances as well as an enhanced role of the competing spin-orbit interaction in materials with heavier anions. At the same time, the dependence of $T_{\mathrm{C}}$ on the effective Mn concentration\cite{MacDonald:2005_NM,Wang:2008_APL} and the density of itinerant holes, changed in (Ga,Mn)As by Mn concentration, donor compensation or by gating \cite{MacDonald:2005_NM,Jungwirth:2006_RMP,Nishitani:2010_PRB}, is consistent with the $p-d$ Zener model. Furthermore, the model describes properly the magnitude of the strain-induced magnetic anisotropy\cite{Glunk:2009_PRB}.

Unfortunately, a more detailed comparison between theory and experiment is hampered by the lack of information on short-range antiferromagnetic interactions, which become progressively more important when the Mn concentration increases, as well as by enduring difficulties in the accurate determination of the Mn and hole densities. According to 3D atom probe investigations\cite{Kodzuka:2009_U}, Mn ions are distributed randomly in (Ga,Mn)As within a 1~nm resolution. Nevertheless, from previous channeling studies\cite{Yu:2002_PRB} we know that because of self-compensation, a considerable portion of Mn ions occupies interstitial positions, and hence they act as double donors\cite{Jungwirth:2006_RMP}, compensating partly holes as well as Mn spins, due to a supposedly antiferromagnetic coupling between interstitial and substitutional Mn pairs. Furthermore, a large fraction of Mn, particularly in annealed samples, reside in a near-to-the-surface MnO layer.

The applicability of the $p-d$ Zener model for (Ga,Mn)As and related systems has been confirmed by {\em ab initio} studies in which inaccuracies of the LSDA are partly compensated by the LSDA + U approach\cite{Wierzbowska:2004_PRB} or by self-interaction corrections\cite{Schulthess:2005_NM}.
Furthermore, a number of ferromagnetism models, tailored to DMSs without holes in the valence band, have been put forward,  as reviewed elsewhere\cite{Jungwirth:2006_RMP,Sato:2010_RMP}. It is still unclear, however, whether a long range ferromagnetic order can settle, say, above 10~K, if holes are bound to individual Mn acceptors in DMSs (the strongly localized regime), so that the exchange interaction decays exponentially with the distance between spin pairs. So-far, a ferromagnetic coupling between isolated nearest neighbor Mn pairs was revealed be scanning tunneling microscopy in GaAs:Mn, and  analyzed successfully in terms of a tight binding model\cite{Kitchen:2006_N}.

It is instructive to compare (Ga,Mn)As, (Ga,Mn)P, and (Ga,Mn)N containing the same concentration of Mn, say, 6\%, as in these three material systems $E_{\mathrm{I}}$ differs significantly, according to the data collected in Fig.~\ref{fig:EI}.  The (Ga,Mn)As films containing 6\% of Mn and typically above  $10^{20}$ holes per cm$^{3}$ are on the metallic side of the MIT\cite{Matsukura:1998_PRB}. In the case of Ga$_{0.94}$Mn$_{0.06}$P the magnitude of $E_{\mathrm{I}}$ is large enough to result in hole localization. However, the magnitude of the conductance activation energy (ref.~\onlinecite{Scarpulla:2005_PRL}), a factor of ten smaller than $E_{\mathrm{I}}$ for GaP:Mn, indicates that the holes are only weakly localised. Accordingly, also in this case the $p-d$ Zener model can serve to explain the origin of ferromagnetic correlations.

\begin{figure}
\includegraphics[width=8.5cm]{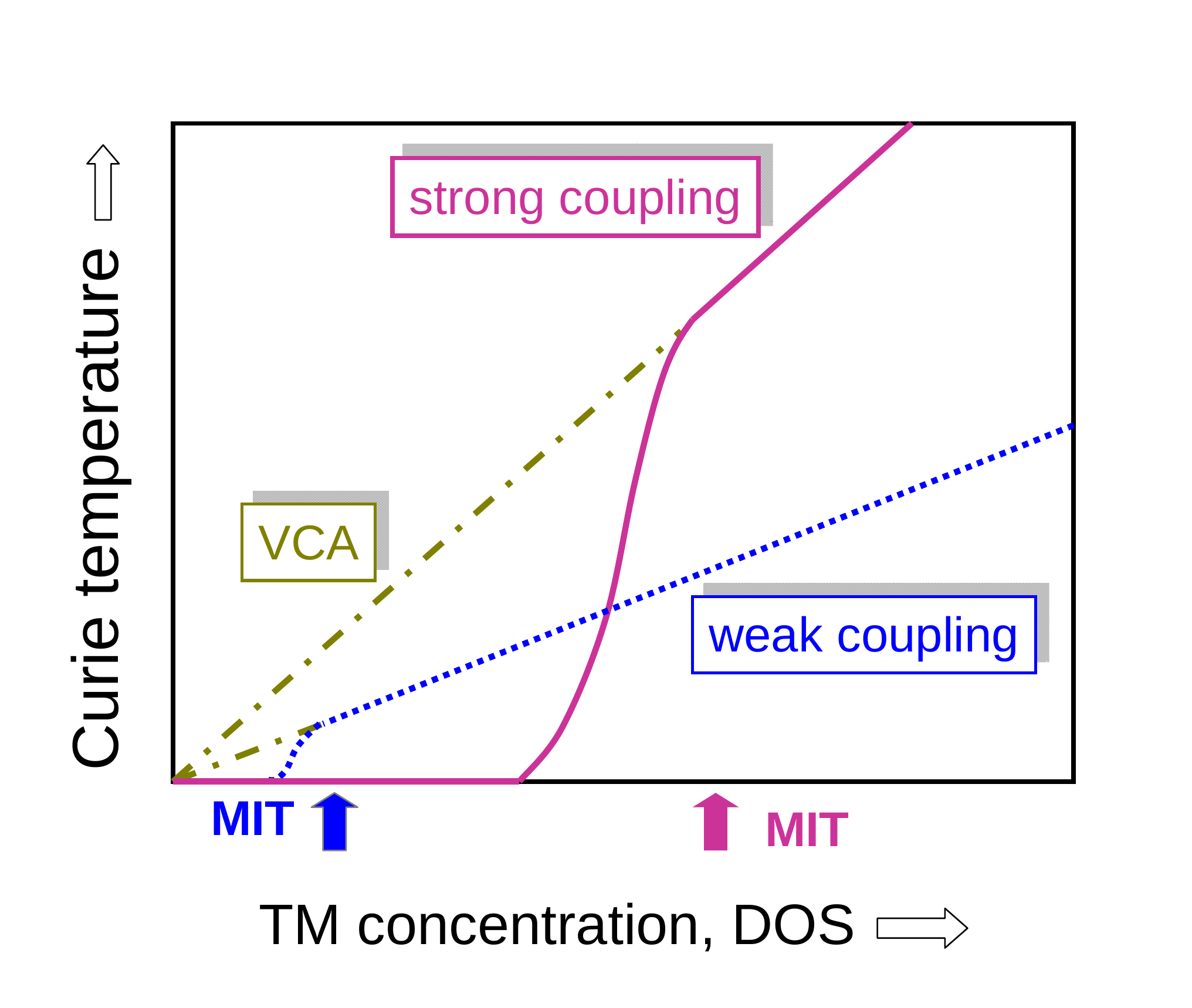}
\caption{Schematic dependence of $T_{\text{C}}$ on the concentration of magnetic impurities and density of hole states at the Fermi level for a weak and a strong coupling. Higher values of $T_{\text{C}}$ are predicted within the virtual crystal and molecular field approximation for the strong coupling. However, the region, where the holes are localized and do not mediate the spin-spin  interaction is wider in the strong coupling case (after~ref.~\onlinecite{Dietl:2008_PRB}).}
\label{fig:VCA_MIT}
\end{figure}

In contrast, no information on hole transport is available for the MBE-grown Ga$_{0.94}$Mn$_{0.06}$N film\cite{Sarigiannidou:2006_PRB}, indicating that the strongly localised regime is reached. In line with the notion that itinerant holes are necessary to observe a coupling between diluted spins, the observed $T_{\mathrm{C}}$ is as low as 8~K. In terms of a schematic drawing presented in Fig.~\ref{fig:VCA_MIT}, (Ga,Mn)N represents the strong coupling case, where the holes remain localised over a wide rage of Mn concentrations, in contrast to both Ga$_{1-x}$Mn$_x$As, for which the MIT appears already below $x = 2\%$ for weakly compensated samples\cite{Jungwirth:2007_PRB}, and (Ga,Mn)P representing an intermediate case.

It is worth noting that there are indications of a non-random Mn distribution in another (Ga,Mn)N film grown by MBE (ref.~\onlinecite{Martinez-Criado:2005_APL}). This may suggest that $T_{\mathrm{C}} = 8$~K constitutes actually an upper limit for $T_{\mathrm{C}}$ in Ga$_{0.94}$Mn$_{0.06}$N. These findings demonstrate, therefore, that the current {\em ab initio} theory\cite{Sato:2010_RMP}, predicting $T_{\mathrm{C}} \approx 60$~K for Ga$_{0.94}$Mn$_{0.06}$N, still overestimates the significance of ferromagnetic couplings in the case of DMSs with no valence band holes.

\section*{COMPETING INTERACTIONS IN {p}-TYPE II-VI {DMSs}}

In II-VI compounds, where Mn is an isoelectronic impurity, it is possible to control independently the spin and the carrier density. However, at given Mn and hole concentrations, $T_{\mathrm{C}}$ is much lower in II-VI DMSs, comparing to III-V compounds, owing to a destructive influence of the short-range antiferromagnetic superexchange.  This effect is less relevant in III-V DMSs, where Mn$^{2+}$ centres are negatively charged, so that the enhanced hole density at closely lying Mn pairs compensates, at least partly, short-range antiferromagnetic interactions\cite{Dietl:2000_S}.

Following theoretical prediction\cite{Dietl:1997_PRB}, carrier-induced ferromagnetism was revealed in modulation doped p-type (Cd,Mn)Te/(Cd,Zn,Mn)Te:Mg heterostructures employing photoluminescence spectroscopy\cite{Haury:1997_PRL,Boukari:2002_PRL}.  The character of magnetic anisotropy as well as the magnitude of $T_{\mathrm{C}}$ and its evolution with the hole density, controlled by the electric field and illumination, were found to be consistent with the $p-d$ Zener model, adapted for this low dimensionality system. Interestingly, however, no magnetic hystereses have been detected below $T_{\mathrm{C}}$.  According to extensive Monte-Carlo simulations, the effect reflects fast magnetization dynamics, generated by the antiferromagnetic interactions at the borders of the hole layer\cite{Lipinska:2009_PRB}.

\begin{figure}
\includegraphics[width=8.5cm]{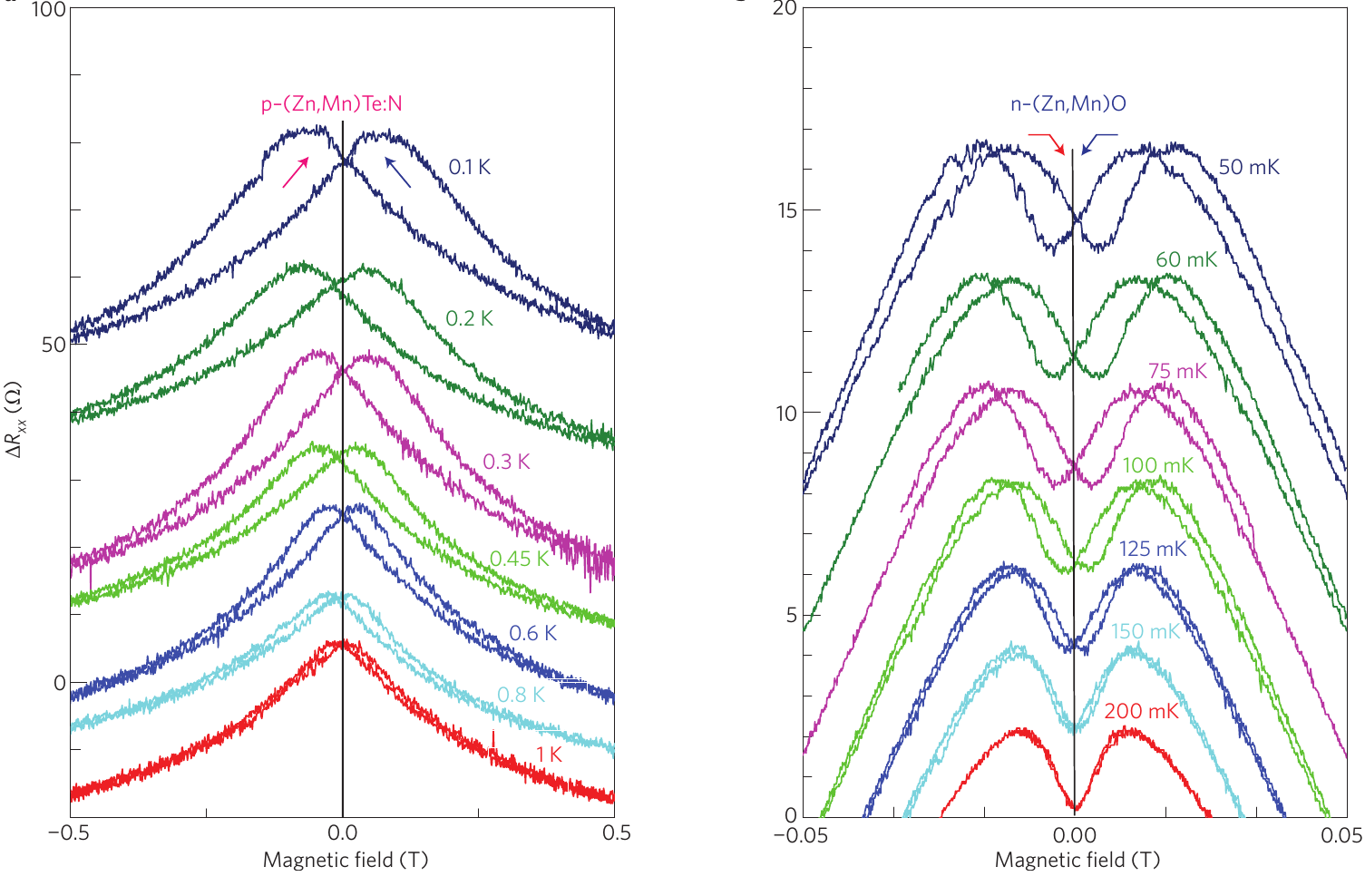}
\caption{Resistive indications of ferromagnetism in p-Zn$_{0.981}$Mn$_{0.019}$Te:N and n-Zn$_{0.97}$Mn$_{0.03}$O:Al.  The temperature dependence of the hystereses width at low temperatures as well as magnetic susceptibility measurements above 2 K point to $T_{\mathrm{C}}  = 1.45\pm 0.05$~K in p-Zn$_{0.981}$Mn$_{0.019}$Te:N with the hole concentration $1.2\cdot 10^{20}$~cm$^{-3}$. The temperature and field scales are an order of magnitude smaller in n-Zn$_{0.97}$Mn$_{0.03}$O:Al with the electron concentration $1.4\cdot 10^{20}$~cm$^{-3}$, where $T_{\mathrm{C}} = 160 \pm 20$~mK (adapted from refs.~\onlinecite{Ferrand:2001_PRB} and \onlinecite{Andrearczyk:2001_ICPS}).}
\label{fig:ZnMnTe_ZnMnO}
\end{figure}

Ferromagnetic signatures were also found in epilayers of p-type (Zn,Mn)Te:N (refs~\onlinecite{Ferrand:2000_JCG, Ferrand:2001_PRB}) and n-type (Zn,Mn)O:Al (ref.~\onlinecite{Andrearczyk:2001_ICPS}). As shown in Fig.~\ref{fig:ZnMnTe_ZnMnO}, under the high density of carriers, the low-temperature resistance acquires a hysteretic behaviour. This points to the appearance of a ferromagnetic order as well as demonstrates directly the existence of a strong coupling between carriers and Mn spins. The temperature dependence of the coercive force, together with magnetic susceptibility measurements above 2~K, point to a magnetic ordering temperature $T_{\mathrm{C}} = 1.45$~K in the case of Zn$_{0.0981}$Mn$_{0.019}$Te:N containing $1.2\cdot 10^{20}$ holes per cm$^{3}$.  This value is in agreement with the predictions of the $p-d$ Zener model, provided that the aforementioned antiferromagnetic interactions and the spin-orbit interaction are taken carefully into account\cite{Ferrand:2001_PRB}. A similar experimental procedure leads to $T_{\mathrm{C}} = 160$~mK in the case of Zn$_{0.097}$Mn$_{0.03}$O:Al containing $1.4\cdot 10^{20}$ electrons per cm$^{3}$ (ref.~\onlinecite{Andrearczyk:2001_ICPS}). Taking the differences in relevant parameters and, in particular, a three times larger amplitude of the $p-d$ exchange integral comparing to the $s-d$ case, the experimentally observed difference in the $T_{\mathrm{C}}$ values between p-type and n-type materials can be readily explained.

\section*{PROSPECTS FOR HIGHER {$T_{\mathrm{C}}$} IN {DMSs}}

A number of authors has reported the observation of room temperature ferromagnetism in various semiconductors and oxides containing supposedly randomly distributed localised spins. However, it is probably fair to say that so-far none of these findings has been confirmed by other groups as well as none has resulted in the demonstration of a device structure working at room temperature. As we will argue in the following sections, if not representing an experimental artifact, this robust ferromagnetism can be explained assuming a {\em non}-random distribution of the magnetic ions.

It appears that despite ten years of extensive investigations the conditions under which the room temperature ferromagnetism was predicted for nitrides and oxides\cite{Dietl:2000_S} have not yet been experimentally met. In particular, no (Ga,Mn)N, (Zn,Mn)O, or a related compound containing a few percent of randomly distributed magnetic cations {\em and} a few $10^{20}$ delocalised or weakly localised holes per cm$^{3}$ has so-far been synthesised. On the other hand, in annealed (Ga,Mn)As the effective Mn concentration, as judged from the magnitude of the saturation magnetization,  approaches now 10\% but $T_{\mathrm{C}}$ remains below 200~K.

Can, therefore, carrier mediated indirect spin-spin coupling produce a ferromagnetic ordering stable up to the  room temperature?  We note that a variant of the $p-d$ Zener model explains the origin of ferromagnetism in double perovskite compounds, such as Sr$_2$CrReO$_6$, where the magnitudes of $T_{\mathrm{C}}$ attain 625~K (ref.~\onlinecite{Serrate:2007_JPCM}), despite that the distance between localized spins is as large as 0.6 -- 0.7~nm. This shows a potential of this mechanism to support  high-temperature ferromagnetism, possibly also in DMSs, where the distance between \textit{second} nearest neighbour cations is smaller than 0.5~nm in GaN and ZnO.

However, the obvious difficulty is to synthesise a p-type wide band gap DMS systems, in which the hole density is large enough to result in a MIT, even in the strong coupling case, as sketched in Fig.~\ref{fig:VCA_MIT}. It has been suggested\cite{Dietl:2008_PRB}---but not yet verified experimentally---that once the MIT is reached, which means that the bound states are washed out by many body screening, high values of $T_{\mathrm{C}}$, expected within the VCA and MFA\cite{Dietl:2000_S}, should emerge. Various methods allowing to enlarge the hole density above $10^{20}$~cm$^{-3}$ without increasing the degree of disorder, such as gating as well as doping in a modulated fashion or by exploiting interfacial electric fields, may constitute the appropriate road towards achieving a semiconductor showing high and tunable $T_{\mathrm{C}}$ values. We note that the enduring progress in the gate oxide deposition\cite{Chiba:2008_N,Sawicki:2010_NP} allows one to achieve an interfacial charge density of the order of $3\cdot 10^{13}$~cm$^{-2}$, that is up to about $3\cdot 10^{20}$ per cm$^{3}$.

\section*{ORIGIN OF HIGH TEMPERATURE FERROMAGNETISM}

Perhaps the most surprising development of the last decade in the science of magnetic materials is abundant observations of spontaneous magnetization persisting up to above room temperature in semiconductors and oxides, in which no ferromagnetism at any temperature has been expected, particularly within the $p-d$ Zener model.  These findings have offered prospects for a spread of spintronic functionalities much wider than it could initially be anticipated. At the same time, they have generated a considerable theoretical effort resulting in proposals of several novel mechanisms of exchange interactions between {\em diluted} spins, designed to interpret robust ferromagnetism in magnetically doped or even magnetically undoped systems. Nevertheless, it appears that there is no visible convergence between particular experimental findings and theoretical models.

Over the last years we start to realize that open $d$ shells of magnetic impurities in non-magnetic solids not only provide localised spins  but $\textit{via}$ hybridisation with band states contribute significantly to the cohesive energy, particularly if TM impurities occupy neighboring sites. The resulting attractive force between magnetic cations leads to their aggregation invalidating the main paradigm of the DMS and DMO  physics concerning the random distribution of TM spins. It may be anticipated that the magnetic nanocrystals formed in this way assume the crystallographic form imposed by the matrix. Accordingly, the  properties of these condensed magnetic semiconductors (CMSs) may be not yet included in materials compendia, so that it is {\em a priori} unknown whether they are metallic or insulating as well as whether they exhibit ferromagnetic, ferrimagnetic or antiferromagnetic spin order. However, due to a large concentration of the magnetic constituent within the CMSs nanocrystals, their spin ordering temperature is expected to be relatively high, typically above the room temperature.

\begin{figure}[ht]
\begin{center}
\includegraphics[width=8.5cm]{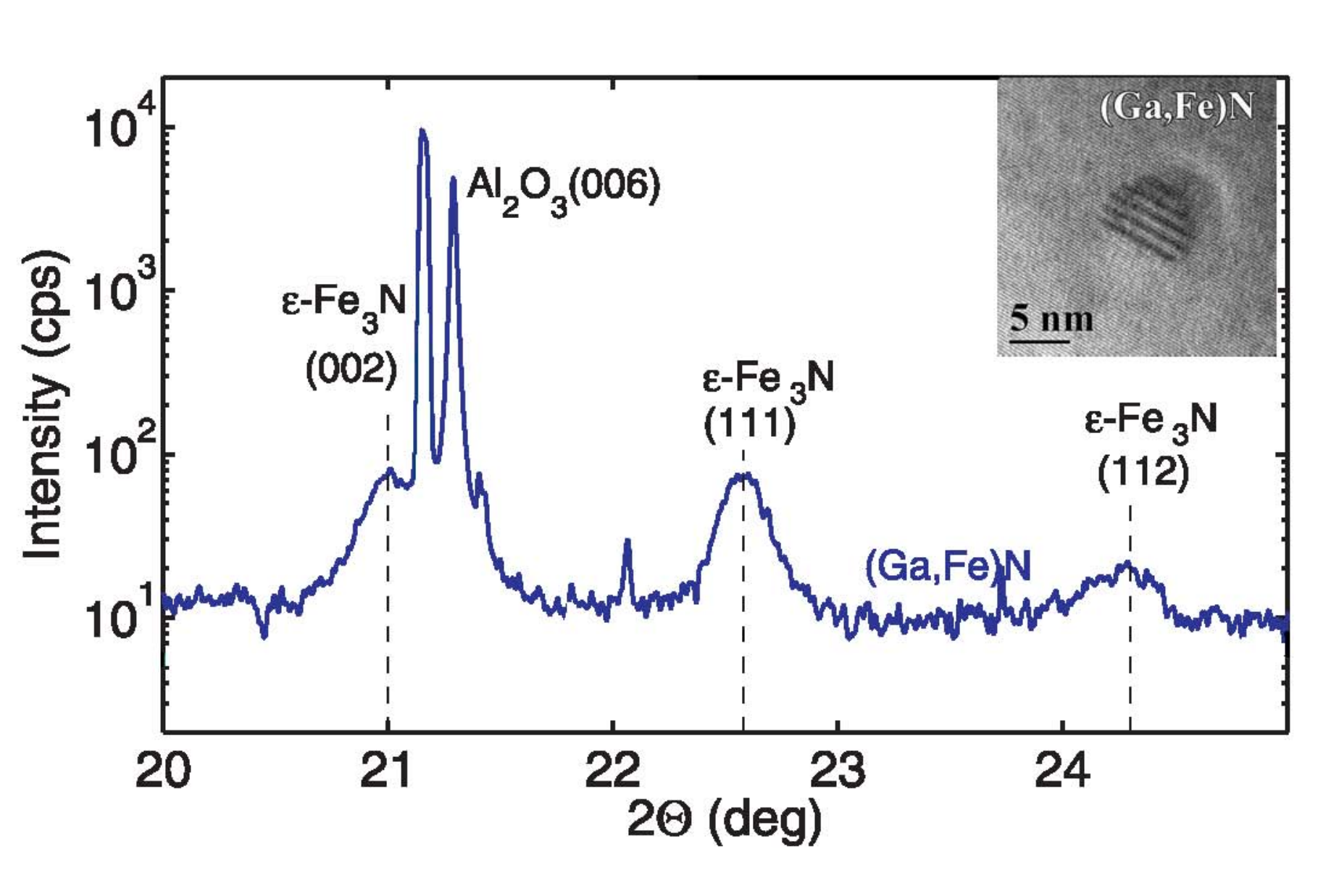}
\includegraphics[width=8.5cm]{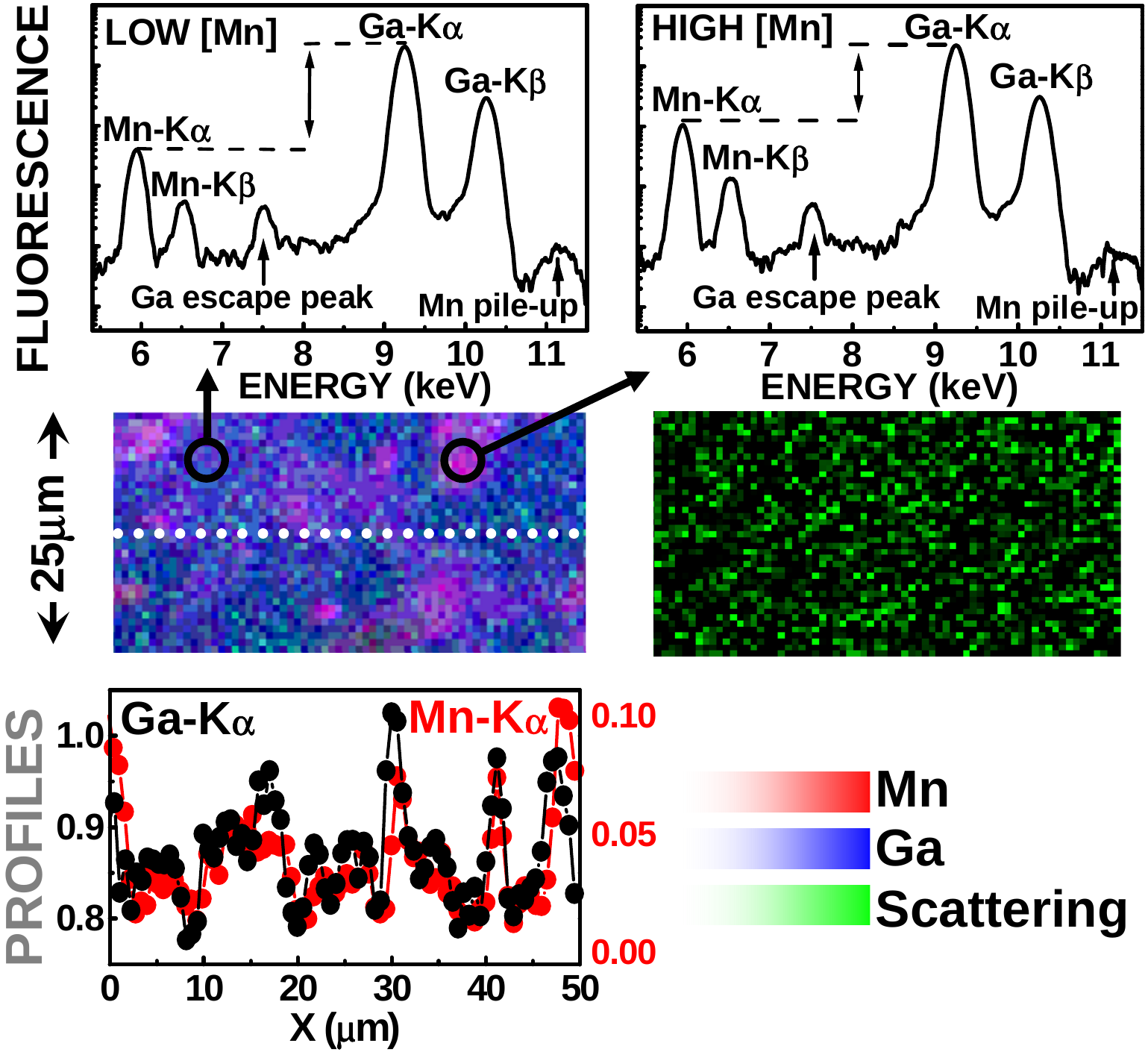}
\caption{Evidence for crystallographic and chemical phase separations in DMSs. Upper panel: synchrotron XRD and TEM results for (Ga,Fe)N showing the precipitation of hexagonal $\epsilon$-Fe$_3$N nanocrystals (after ref.~\onlinecite{Bonanni:2008_PRL}). Lower panel: element-specific synchrotron radiation micro-probe analysis of (Ga,Mn)N showing aggregation of Mn cations (after ref.~\onlinecite{Martinez-Criado:2005_APL}).}
\label{fig:GaFeN_GaMnN}
\end{center}
\end{figure}

As seen today, the experimental detection of a non-random spin distribution and possible contamination has been highly challenging in DMSs research\cite{Bonanni:2007_SST}. Only recently the actual spatial distribution of TM cations in some DMSs has been established by some groups, owing to the application of state-of-the-art element-specific nanocharacterisation tools. While in some cases one deals with elemental ferromagnetic metal nanoparticles, the case of Co in ZnO (ref.~\onlinecite{Park:2004_APL}), usually TM compounds are involved. Taking (Ga,Fe)N as an example, we note that according to standard laboratory high-resolution x-ray diffraction (HRXRD), the incorporation of Fe simply leads to a broadening of the GaN-related diffraction maxima without revealing any secondary phases\cite{Bonanni:2007_PRB}. In contrast, a much brighter synchrotron source has allowed to identify the presence of precipitates in the same samples, as shown in Fig.~\ref{fig:GaFeN_GaMnN}, a counterpart of MnAs nanocrystals in GaAs (ref.~\onlinecite{De_Boeck:1996_APL,Moreno:2002_JAP}). The appearance of crystallographic phase separation in (Ga,Fe)N is supported by near-edge x-ray absorption fine-structure (EXAFS) studies\cite{Rovezzi:2009_PRB}. The dominant ferromagnetic precipitate was identified as Fe$_3$N but in some cases nanocrystals in the form of an elemental ferromagnetic metal, Fe in this case, are also visible\cite{Bonanni:2008_PRL}. At the same time, transmission electron microscopy (TEM) with appropriate mass and strain contrast as well as electron dispersive spectroscopy (EDS), not only corroborated the outcome of synchrotron XDR, but revealed also the aggregation of magnetic cations without distorting the host wurtzite structure under certain growth conditions\cite{Bonanni:2008_PRL}.  This chemical phase separation is known in the DMS literature as spinodal decomposition, independently of the microscopic mechanism leading to the aggregation of the TM cations.

\begin{figure}[ht]
\begin{center}
\includegraphics[width=8.7cm]{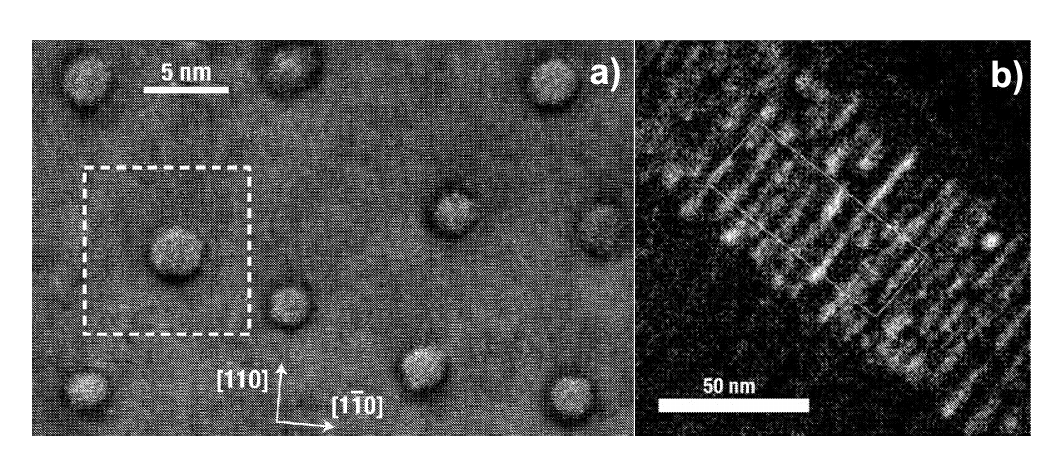}
\includegraphics[width=8.5cm]{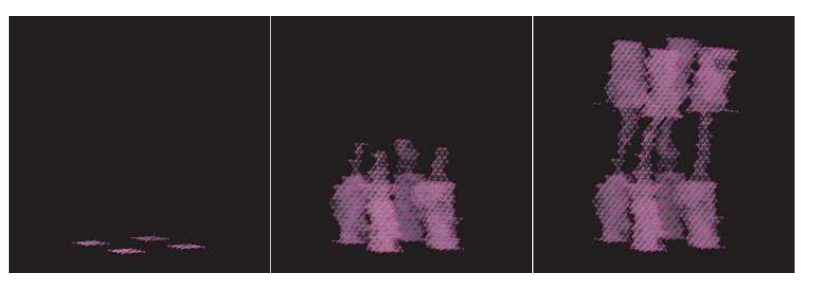}
\caption{Formation of nanocolumns in DMS by aggregation of Mn cations. Upper panel: Mn-rich nanocolumns in (Ge,Mn) evidenced by: (a) HRTEM plane view and (b) Mn chemical maps (after ref.~\onlinecite{Jamet:2006_NM}). Lower panel: Monte Carlo simulation of chemical phase separation in (Zn,Cr)Te with seeding to initiate the growth of nanocolumns and control of their diameter by Cr flux (after ref.~\onlinecite{Katayama-Yoshida:2007_PSSA}).}
\label{fig:GeMn_HKY}
\end{center}
\end{figure}

The application of TEM with EDS allowed to evidence the chemical phase separation in annealed (Ga,Mn)As (ref.~\onlinecite{Moreno:2002_JAP,Tanaka:2008_B}), (Zn,Cr)Te (ref.~\onlinecite{Kuroda:2007_NM}), (Al,Cr)N, and (Ga,Cr)N (ref.~\onlinecite{Gu:2005_JMMM}), whereas according to the results summarised in Fig.~\ref{fig:GaFeN_GaMnN}, hexagonal nanocrystals were detected in (Ga,Mn)N. Finally, we mention the case of (Ge,Mn), where under suitable growth conditions {\em quasi}-periodically arranged nanocolumns are observed\cite{Jamet:2006_NM}, as shown in Fig.~\ref{fig:GeMn_HKY}. Actually, a tendency to nanocolumn formation was also reported for (Al,Cr)N (ref.~\onlinecite{Gu:2005_JMMM}) and (Zn,Cr)Te (ref.~\onlinecite{Nishio:2009_MRSP}).  This demonstrates that growth conditions can assist in controlling the nanocrystals shape. Interestingly, these two kinds of nanocrystal forms were reproduced by Monte-Carlo simulations\cite{Katayama-Yoshida:2007_PSSA}. Furthermore, a strict correlation between ferromagnetic features and the presence of CMS nanocrystals has been demonstrated for these systems. However, the identification of the dominant microscopic mechanisms leading to robust spin ordering, that is to a large magnitude of $T_{\mathrm{C}}$ and magnetic anisotropy, awaits for detailed experimental and theoretical studies for particular combinations of CMSs and hosts.

As expected, the lowering of the growth temperature and/or the increase of the growth rate\cite{Bonanni:2008_PRL} hampers the aggregation of magnetic cations.  Moreover, it has been suggested that it is possible to change the TM charge state and, therefore, the aggregation energy by co-doping with shallow donors or acceptors\cite{Dietl:2006_NM,Ye:2006_PRB}. This way of affecting the TM valency stems from the presence of band gap states derived from the $d$ orbitals. These states trap carriers supplied by shallow impurities, altering the charge state of the magnetic cations and, hence, modifying their mutual interactions.  Accordingly, co-doping of DMSs and DMOs with shallow acceptors or donors, during either growth or post-growth processing, modifies the valence providing a powerful mean for the control of the magnetic cation aggregation. These predictions are corroborated by experimental finding for (Ga,Mn)N (ref.~\onlinecite{Bonanni:2007_SST}), (Zn,Cr)Te (ref.~\onlinecite{Kuroda:2007_NM}), and (Ga,Fe)N (ref.~\onlinecite{Bonanni:2008_PRL}), where remarkable changes in the ferromagnetic characteristics upon co-doping with shallow impurities have been found and correlated with the TM distribution.

Finally, we note that depending on the growth conditions CMS nanocrystals can be distributed randomly or accumulate either at the interface with the buffer or at the film surface. Furthermore, TM impurities may decorate or diffuse along extended defects such as dislocations or grain boundaries\cite{Straumal:2009_PRB}. This appears to explain an inverse correlation between samples' quality and the appearance of high $T_{\mathrm{C}}$ ferromagnetism, noted by some authors in the case of oxides\cite{Ney:2008_PRL}.

\section*{SOME MODELING}

The above qualitative considerations are supported by {\em ab initio} studies. For instance, the computed energy change associated with bringing two Ga-substitutional Mn atoms to the nearest neighbor cation position is $E_{\mathrm{d}} = -120$~meV in GaAs and $-300$~meV in GaN, and reaches $-140$ and $-350$~meV in the case of a cation-substitutional Cr nearest neighbor pair in ZnTe and GaN, respectively\cite{Schilfgaarde:2001_PRB,Kuroda:2007_NM}. In contrast,  there is virtually  no energy change associated with bringing two Zn-substitutional Mn atoms to the nearest neighbor cation sites in (Zn,Mn)Te, where $E_{\mathrm{d}} = 21$~meV (ref.~\onlinecite{Kuroda:2007_NM}). This can be associated to the fact that the Mn $d$ states little perturb the $sp^3$ tetrahedral bonds as both the lower $d^5$ (donor) and the upper $d^6$ (acceptor) Hubbard levels are respectively well below and above the band edges in II-VI compounds\cite{Zunger:1986_SSP}, so that there is no considerable difference between the band hybridisation involving Zn or Mn. This conclusion is consistent with a large solubility of Mn in II-VI compounds and the apparently random distribution of Mn in these systems\cite{Furdyna:1988_B}.

\begin{figure}
\includegraphics[width=9cm]{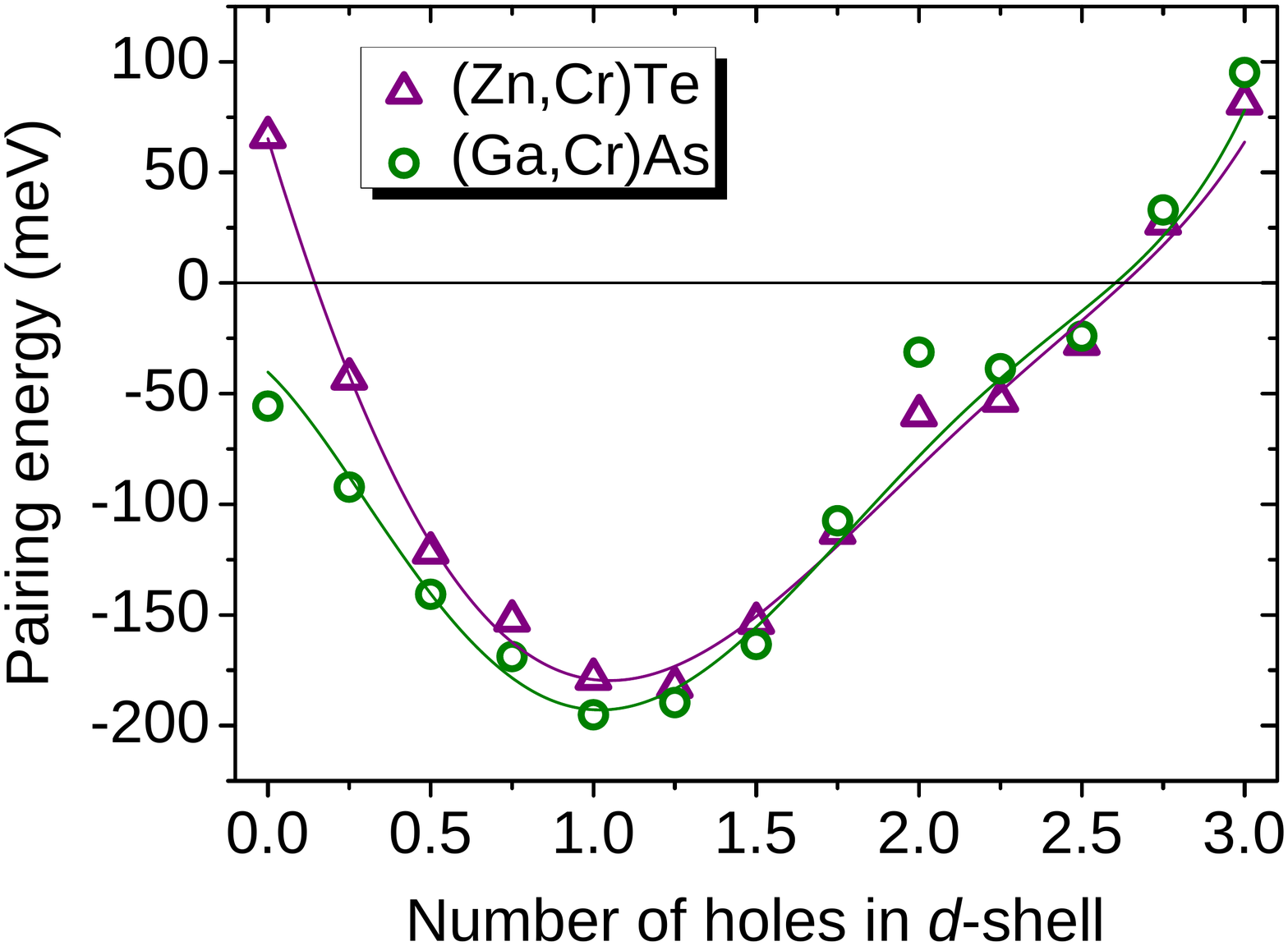}
\caption{Computed energy change $E_{\text{d}}$ resulting from bringing two Cr impurities to the nearest neighbor cation positions in ZnTe and GaAs depending on the number of holes in the Cr $d^5$ shell  (adapted from ref.~\onlinecite{Da_Silva:2008_NJP}).}
\label{fig:ZnCrTe_Da_Silva}
\end{figure}

To model the effect of co-doping, we note that the energy of the {\em screened} Coulomb interaction between two elementary charges residing on the nearest neighbor cation sites in the GaAs lattice is $E_{\mathrm{d}}=-280$~meV. This value indicates that a change in the charge state can affect the aggregation significantly, as the gain of energy associated with bringing two Mn atoms close by is $E_{\mathrm{d}} = -120$~meV, as quoted above\cite{Schilfgaarde:2001_PRB}.  Accordingly, a surplus of charge on TM ions, comparing to non-magnetic cations, brought by co-doping with shallow dopants can overweight the gain of energy stemming from $p-d$ hybridization and impede the nanocrystal assembling\cite{Dietl:2006_NM,Ye:2006_PRB}. This picture is confirmed by \emph{ab initio} computations within the LSDA for (Ti,Cr)O$_2$ (ref.~\onlinecite{Ye:2006_PRB}) and (Zn,Cr)Te (refs~\onlinecite{Kuroda:2007_NM} and \onlinecite{Da_Silva:2008_NJP}). As shown in Fig.~\ref{fig:ZnCrTe_Da_Silva}, the value of $E_{\mathrm{d}}$ attains a minimum in ZnTe when the two Cr cations are in the $2+$ charge state\cite{Da_Silva:2008_NJP} ($d^4$ configuration). However, the computation results shown in the same plot indicate that also in GaAs, $E_{\mathrm{d}}$ goes through a minimum for the Cr$^{2+}$ case, rather than in the case of Cr$^{3+}$ pairs, as might be expected for III-V compounds.

\section*{FROM DILUTE TO NANOCOMPOSITE SYSTEMS}

In view of the above discussion, the incorporation of TM impurities to semiconductors not only bridges ferromagnetic and semiconductor capabilities but also offers a way to develop a new kind of nanocomposite systems consisting of ferromagnetic and metallic nanocrystals coherently buried in a semiconductor host.  The application of embedded metallic and semiconducting nanocrystals is known to be on the way to revolutionise the performance of various commercial devices, such as flash memories, low current semiconductor lasers, and single photon emitters. Similarly far reaching can be the use of nanocomposite semiconductor/ferromagnetic systems due to their unique capabilities and to the possibility of controlling the shape (nanodots {\em vs.} nanocolumns) and size by growth parameters and co-doping during the epitaxial process.

It has already been demonstrated that these nonocomposites show strong magnetotrasport and magnetooptical effects\cite{Kuroda:2007_NM,Tanaka:2008_B}, that could possibly allow these systems to be exploited as magnetic field sensors as well as in magnetooptical devices. In particular, a combination of a strong magnetic circular dichroism specific to ferromagnetic metals and weak losses characterising the semiconductor hosts suggest possible functionalities as optical isolators as well as three-dimensional (3D) tunable photonic crystals and spatial light modulators for advanced photonic applications. As an interesting recent development one can quote a spin battery effect demonstrated for the GaAs:MnAs system in a magnetic field\cite{Hai:2009_N}.

As shown in Fig.~\ref{fig:GeMn_HKY}, the controlled growth of nanocolumns of a ferromagnetic metal can allow to fabricate {\em in-situ}, {\em e.g.}, a dense array of magnetic tunnel junctions\cite{Katayama-Yoshida:2007_PSSA} or Coulomb blockade devices. Thus, the media in question can be employed for low-power high-density magnetic storage, including spin-torque magnetic random access memories and race-track 3D domain-wall based memories.  If sufficiently high tunneling magnetoresistance (TMR) will be found, one can envisage the application for field programmable logic (TMR-based connecting/disconnecting switches) and even for all-magnetic logic, characterised by low power consumption and radiation hardness. Furthermore, embedded metallic nanostructures may serve as building blocks for all-metallic nanoelectronics and for high quality nanocontacts in nanoelectronics,  optoelectronics, and plasmonics as well as constitute media for thermoelectric applications\cite{Katayama-Yoshida:2008_B}. Worth mentioning is also the importance of hybrid semiconductor/ferromagnetic systems in various proposals of scalable quantum processors.

\section*{$d^{0}$ FERROMAGNETISM AND BEYOND}

Organic ferromagnets and quantum Hall ferromagnets are a proof that ferromagnetism is possible in materials without magnetic ions, albeit the corresponding $T_{\mathrm{C}}$'s are so-far rather low, typically below 20~K. It has also been suggested that a robust ferromagnetism can appear in certain zinc-blende metals like CaAs, and be driven by a Stoner instability in the narrow heavy hole band\cite{Geshi:2005_AIPCP,Volnianska:2010_JPCM}, a prediction awaiting for an experimental confirmation.

It has been known for a long time that a number of defects or non-magnetic impurities form localised paramagnetic centers in various hosts. Some of these states might show a large intra-center correlation energy $U$ that could ensure an adequate stability of the spins, even if their density increases or the material is co-doped with shallow impurities. It is, therefore, tempting to relate the presence of unexpected high-temperature ferromagnetism in various oxides and carbon derivates to magnetic moments residing rather on nonmagnetic defects or impurities than on open $d$ shells of TMs\cite{Elfimov:2002_PRL,Coey:2006_COSSMS}.

As shown theoretically\cite{Sliwa:2008_PRB}, a sizable exchange interaction takes place between valence-band holes residing on non-magnetic acceptors and electrons in the conduction band. This demonstrates that band carriers can mediate a Zener-type coupling between spins localised on defect centers. Furthermore, if the spin concentration increases, so that either the Hubbard-Mott or the Anderson-Mott transition is reached, the double exchange or Stoner-like mechanism might appear\cite{Coey:2008_JPD}. However, in each of these cases a clear correlation between magnetic and transport properties should be visible, analogous to the one observed routinely in manganites, (Ga,Mn)As, and $p$-(Zn,Mn)Te.  In particular, the fabrication of a spintronic structure -- like a magnetic tunnel junction --  working up to high temperatures, would constitute a strong confirmation of the existence of spin transport in these challenging systems. Alternatively, defects and impurities, similarly to TM dopants, could form high spin aggregates in certain hosts. In this case, the presence of ferromagnetic-like features should correlate with the existence of defect agglomerations that can be revealed by employing state-of-the-art nanocharacterisation tools.

To date, suggestions concerning defect-related high-temperature ferromagnetism come only from global magnetization measurements. Therefore, it appears more natural to assume at this stage that a small number of magnetic nanoparticles -- that escaped from the detection procedure -- account for the high-temperature ferromagnetic-like behaviour of nominally nonmagnetic insulators and semiconductors. Such nanoparticles could be introduced during the synthesis or post-growth processing, and can reside in the sample volume, at dislocations or grain boundaries but also at the surface, interface or in the substrate. An instructive example is provided by the case of porous silicon\cite{Grace:2009_AM}.

\section*{SUMMARY}

Where are we then after these ten years? With no doubt (Ga,Mn)As and related compounds have considerably strengthen their position as an outstanding playground to develop and test novel functionalities unique to a combination of ferromagnetic and semiconductor systems. Many of the concepts\cite{Spintronics:2008_B}, like spin-injection, electric-field control of the $T_{\mathrm{C}}$ magnitude and magnetization direction, tunneling anisotropic magnetoresistance in planar junctions and in the Coulomb blockade regime, current-induced domain displacement without assistance of a magnetic field, are being now developed in devices involving ferromagnetic metals, which may function at ambient temperatures. Obviously, however, a further increase of $T_{\mathrm{C}}$, over the current record value of 190~K, continues to be a major goal in the field of DMSs.

At the same time, investigations of magnetically doped semiconductors and oxides have faced us with unexpectedly challenging issues, intractable by conventional materials characterisation, computational, and theoretical tools.  In addition to an interplay between phenomena specific to strongly correlated and disordered systems, encountered in doped semiconductors and manganites, DMSs and DMOs show a fundamentally new ingredient brought about by a not anticipated correlation in the magnetic ion distribution.  We are now learning how to visualise and control the magnetic ion aggregation in order to develop methods allowing to obtain lateral and vertical distributions as well as shapes of magnetic nanocrystals on demand. The ferromagnetic metal/semiconductor nanocomposites fabricated in this way offer a spectrum of so-far unexplored possibilities in various fields of materials science and device physics.

Are there only two classes of magnetically doped semiconductors and oxides showing ferromagnetic features, say, above 10~K? Are ferromagnetic correlation mediated by valence band holes and embedded magnetic nanocrystals the only sources of spontaneous magnetization in these systems? We will see in the years to come.

\section*{Acknowledgments}
The author's works described here were supported by the FunDMS Advanced Grant of the European Research Council within the "Ideas" 7th Framework Programme of the EC, and earlier by ERATO project of Japan Science and Technology Agency and Humboldt Foundation. The fruitful and enjoyable collaboration with the groups of Alberta Bonanni, Shinji Kuroda, Hideo Ohno, and Maciej Sawicki is gratefully acknowledged.


\begin{thebibliography}{150}

\expandafter\ifx\csname url\endcsname\relax
  \def\url#1{\texttt{#1}}\fi
\expandafter\ifx\csname urlprefix\endcsname\relax\def\urlprefix{URL }\fi
\providecommand{\bibinfo}[2]{#2}
\providecommand{\eprint}[2][]{\url{#2}}

\bibitem{Story:1986_PRL}
\bibinfo{author}{Story, T.}, \bibinfo{author}{Galazka, R.~R.},
  \bibinfo{author}{Frankel, R.~B.} \& \bibinfo{author}{Wolff, P.~A.}
\newblock \bibinfo{title}{Carrier-concentration--induced ferromagnetism in
  {PbSnMnTe}}.
\newblock \emph{\bibinfo{journal}{Phys. Rev. Lett.}}
  \textbf{\bibinfo{volume}{56}}, \bibinfo{pages}{777--779}
  (\bibinfo{year}{1986}).

\bibitem{Ohno:1992_PRL}
\bibinfo{author}{Ohno, H.}, \bibinfo{author}{Munekata, H.},
  \bibinfo{author}{Penney, T.}, \bibinfo{author}{von {Moln{\'a}r}, S.} \&
  \bibinfo{author}{Chang, L.~L.}
\newblock \bibinfo{title}{Magnetotransport properties of p-type {(In,Mn)As}
  diluted magnetic {III-V} semiconductors}.
\newblock \emph{\bibinfo{journal}{Phys. Rev. Lett.}}
  \textbf{\bibinfo{volume}{68}}, \bibinfo{pages}{2664--2667}
  (\bibinfo{year}{1992}).

\bibitem{Ohno:1996_APL}
\bibinfo{author}{Ohno, H.} \emph{et~al.}
\newblock \bibinfo{title}{{(Ga,Mn)As: A new diluted magnetic semiconductor
  based on GaAs}}.
\newblock \emph{\bibinfo{journal}{Appl. Phys. Lett.}}
  \textbf{\bibinfo{volume}{69}}, \bibinfo{pages}{363} (\bibinfo{year}{1996}).

\bibitem{Esch:1997_PRB}
\bibinfo{author}{Van~Esch, A.} \emph{et~al.}
\newblock \bibinfo{title}{Interplay between the magnetic and transport
  properties in the {III-V} diluted magnetic semiconductor
  {Ga$_{1-x}$Mn$_x$As}}.
\newblock \emph{\bibinfo{journal}{Phys. Rev. B}} \textbf{\bibinfo{volume}{56}},
  \bibinfo{pages}{13103--13112} (\bibinfo{year}{1997}).

\bibitem{Haury:1997_PRL}
\bibinfo{author}{Haury, A.} \emph{et~al.}
\newblock \bibinfo{title}{{Observation of a Ferromagnetic Transition Induced by
  Two-Dimensional Hole Gas in Modulation-Doped CdMnTe Quantum Wells}}.
\newblock \emph{\bibinfo{journal}{Phys. Rev. Lett.}}
  \textbf{\bibinfo{volume}{79}}, \bibinfo{pages}{511--514}
  (\bibinfo{year}{1997}).

\bibitem{Ferrand:2000_JCG}
\bibinfo{author}{Ferrand, D.} \emph{et~al.}
\newblock \bibinfo{title}{{Carrier-induced ferromagnetic interactions in
  p-doped {Zn$_{1-x}$Mn$_{x}$Te} epilayers}}.
\newblock \emph{\bibinfo{journal}{J. Cryst. Growth}}
  \textbf{\bibinfo{volume}{214-215}}, \bibinfo{pages}{387--390}
  (\bibinfo{year}{2000}).

\bibitem{Awschalom:2007_NP}
\bibinfo{author}{Awschalom, D.~D.} \& \bibinfo{author}{{Flatt\'e}, M.~E.}
\newblock \bibinfo{title}{Challenges for semiconductor spintronics}.
\newblock \emph{\bibinfo{journal}{Nature Phys.}} \textbf{\bibinfo{volume}{3}},
  \bibinfo{pages}{153--159} (\bibinfo{year}{2007}).

\bibitem{Spintronics:2008_B}
\bibinfo{editor}{Dietl, T.}, \bibinfo{editor}{Awschalom, D.~D.},
  \bibinfo{editor}{Kaminska, M.} \& \bibinfo{editor}{Ohno, H.} (eds.)
  \emph{\bibinfo{title}{Spintronics}}, vol.~\bibinfo{volume}{82} of
  \emph{\bibinfo{series}{Semiconductors and Semimetals}}
  (\bibinfo{publisher}{Elsevier}, \bibinfo{address}{Amsterdam},
  \bibinfo{year}{2008}).

\bibitem{Ohno:1999_N}
\bibinfo{author}{Ohno, Y.} \emph{et~al.}
\newblock \bibinfo{title}{{Electrical spin injection in a ferromagnetic
  semiconductor heterostructure}}.
\newblock \emph{\bibinfo{journal}{Nature}} \textbf{\bibinfo{volume}{402}},
  \bibinfo{pages}{790--792} (\bibinfo{year}{1999}).

\bibitem{Ohno:2000_N}
\bibinfo{author}{Ohno, H.} \emph{et~al.}
\newblock \bibinfo{title}{{Electric-field control of ferromagnetism}}.
\newblock \emph{\bibinfo{journal}{Nature}} \textbf{\bibinfo{volume}{408}},
  \bibinfo{pages}{944--946} (\bibinfo{year}{2000}).

\bibitem{Chiba:2008_N}
\bibinfo{author}{Chiba, D.} \emph{et~al.}
\newblock \bibinfo{title}{Magnetization vector manipulation by electric
  fields}.
\newblock \emph{\bibinfo{journal}{Nature}} \textbf{\bibinfo{volume}{455}},
  \bibinfo{pages}{515--518} (\bibinfo{year}{2008}).

\bibitem{Chernyshov:2009_NP}
\bibinfo{author}{Chernyshov, A.} \emph{et~al.}
\newblock \bibinfo{title}{Evidence for reversible control of magnetization in a
  ferromagnetic material by means of spin–-orbit magnetic field}.
\newblock \emph{\bibinfo{journal}{Nature Phys.}} \textbf{\bibinfo{volume}{5}},
  \bibinfo{pages}{656--659} (\bibinfo{year}{2009}).

\bibitem{Gould:2004_PRL}
\bibinfo{author}{Gould, C.} \emph{et~al.}
\newblock \bibinfo{title}{{Tunneling Anisotropic Magnetoresistance: A
  spin-valve like tunnel magnetoresistance using a single magnetic layer}}.
\newblock \emph{\bibinfo{journal}{Phys. Rev. Lett.}}
  \textbf{\bibinfo{volume}{93}}, \bibinfo{pages}{117203}
  (\bibinfo{year}{2004}).

\bibitem{Wunderlich:2006_PRL}
\bibinfo{author}{Wunderlich, J.} \emph{et~al.}
\newblock \bibinfo{title}{{Coulomb Blockade Anisotropic Magnetoresistance
  Effect in a (Ga,Mn)As Single-Electron Transistor}}.
\newblock \emph{\bibinfo{journal}{Phys. Rev. Lett.}}
  \textbf{\bibinfo{volume}{97}}, \bibinfo{pages}{077201}
  (\bibinfo{year}{2006}).

\bibitem{Yamanouchi:2004_N}
\bibinfo{author}{Yamanouchi, M.}, \bibinfo{author}{Chiba, D.},
  \bibinfo{author}{Matsukura, F.} \& \bibinfo{author}{Ohno, H.}
\newblock \bibinfo{title}{{Current-induced domain-wall switching in a
  ferromagnetic semiconductor structure}}.
\newblock \emph{\bibinfo{journal}{Nature}} \textbf{\bibinfo{volume}{428}},
  \bibinfo{pages}{539--541} (\bibinfo{year}{2004}).

\bibitem{Matsumoto:2001_S}
\bibinfo{author}{Matsumoto, Y.} \emph{et~al.}
\newblock \bibinfo{title}{Room temperature ferromagnetism in transparent
  transition metal-doped titanium dioxide}.
\newblock \emph{\bibinfo{journal}{Science}} \textbf{\bibinfo{volume}{291}},
  \bibinfo{pages}{854--856} (\bibinfo{year}{2001}).

\bibitem{Bonanni:2010_CSR}
\bibinfo{author}{Bonanni, A.} \& \bibinfo{author}{Dietl, T.}
\newblock \bibinfo{title}{A story of high-temperature ferromagnetism in
  semiconductors}.
\newblock \emph{\bibinfo{journal}{Chem. Soc. Rev.}}
  \textbf{\bibinfo{volume}{39}}, \bibinfo{pages}{528–539}
  (\bibinfo{year}{2010}).

\bibitem{Sheu:2007_PRL}
\bibinfo{author}{Sheu, B.~L.} \emph{et~al.}
\newblock \bibinfo{title}{Onset of ferromagnetism in low-doped
  {Ga$_{1-x}$Mn$_x$As}}.
\newblock \emph{\bibinfo{journal}{Phys. Rev. Lett.}}
  \textbf{\bibinfo{volume}{99}}, \bibinfo{pages}{227205}
  (\bibinfo{year}{2007}).

\bibitem{Sawicki:2010_NP}
\bibinfo{author}{Sawicki, M.} \emph{et~al.}
\newblock \bibinfo{title}{Experimental probing of the interplay between
  ferromagnetism and localization in {(Ga, Mn)As}}.
\newblock \emph{\bibinfo{journal}{Nature Phys.}} \textbf{\bibinfo{volume}{6}},
  \bibinfo{pages}{22} (\bibinfo{year}{2010}).

\bibitem{Richardella:2010_S}
\bibinfo{author}{Richardella, A.} \emph{et~al.}
\newblock \bibinfo{title}{Visualizing critical correlations near the
  metal-insulator transition in {Ga$_{1-x}$Mn$_x$As}}.
\newblock \emph{\bibinfo{journal}{Science}} \textbf{\bibinfo{volume}{327}},
  \bibinfo{pages}{665} (\bibinfo{year}{2010}).

\bibitem{Dietl:2000_S}
\bibinfo{author}{Dietl, T.}, \bibinfo{author}{Ohno, H.},
  \bibinfo{author}{Matsukura, F.}, \bibinfo{author}{Cibert, J.} \&
  \bibinfo{author}{Ferrand, D.}
\newblock \bibinfo{title}{Zener {Model} {Description} of {Ferromagnetism} in
  {Zinc-Blende} {Magnetic} {Semiconductors}}.
\newblock \emph{\bibinfo{journal}{Science}} \textbf{\bibinfo{volume}{287}},
  \bibinfo{pages}{1019--1022} (\bibinfo{year}{2000}).

\bibitem{Matsukura:1998_PRB}
\bibinfo{author}{Matsukura, F.}, \bibinfo{author}{Ohno, H.},
  \bibinfo{author}{Shen, A.} \& \bibinfo{author}{Sugawara, Y.}
\newblock \bibinfo{title}{Transport properties and origin of ferromagnetism in
  {(Ga,Mn)As}}.
\newblock \emph{\bibinfo{journal}{Phys. Rev. B}} \textbf{\bibinfo{volume}{57}},
  \bibinfo{pages}{R2037--R2040} (\bibinfo{year}{1998}).

\bibitem{Okabayashi:1998_PRB}
\bibinfo{author}{Okabayashi, J.} \emph{et~al.}
\newblock \bibinfo{title}{Core-level photoemission study of
  {Ga$_{1-x}$Mn$_{x}$As}}.
\newblock \emph{\bibinfo{journal}{Phys. Rev. B}} \textbf{\bibinfo{volume}{58}},
  \bibinfo{pages}{R4211--R4214} (\bibinfo{year}{1998}).

\bibitem{Linnarsson:1997_PRB}
\bibinfo{author}{Linnarsson, M.}, \bibinfo{author}{{Janz{\'e}n}, E.},
  \bibinfo{author}{Monemar, B.}, \bibinfo{author}{Kleverman, M.} \&
  \bibinfo{author}{Thilderkvist, A.}
\newblock \bibinfo{title}{Electronic structure of the {GaAs:Mn$_{Ga}$} center}.
\newblock \emph{\bibinfo{journal}{Phys. Rev. B}} \textbf{\bibinfo{volume}{55}},
  \bibinfo{pages}{6938--6944} (\bibinfo{year}{1997}).

\bibitem{Zener:1951_PR}
\bibinfo{author}{Zener, C.}
\newblock \bibinfo{title}{Interaction between the d shells in the transition
  metals}.
\newblock \emph{\bibinfo{journal}{Phys. Rev.}} \textbf{\bibinfo{volume}{81}},
  \bibinfo{pages}{440--444} (\bibinfo{year}{1951}).

\bibitem{Dietl:1997_PRB}
\bibinfo{author}{Dietl, T.}, \bibinfo{author}{Haury, A.} \&
  \bibinfo{author}{d'Aubigne, Y.~M.}
\newblock \bibinfo{title}{Free carrier-induced ferromagnetism in structures of
  diluted magnetic semiconductors}.
\newblock \emph{\bibinfo{journal}{Phys. Rev. B}} \textbf{\bibinfo{volume}{55}},
  \bibinfo{pages}{R3347--R3350} (\bibinfo{year}{1997}).

\bibitem{Jungwirth:1999_PRB}
\bibinfo{author}{Jungwirth, T.}, \bibinfo{author}{Atkinson, W.~A.},
  \bibinfo{author}{Lee, B.} \& \bibinfo{author}{MacDonald, A.~H.}
\newblock \bibinfo{title}{{Interlayer coupling in ferromagnetic semiconductor
  superlattices}}.
\newblock \emph{\bibinfo{journal}{Phys. Rev. B}} \textbf{\bibinfo{volume}{59}},
  \bibinfo{pages}{9818--9821} (\bibinfo{year}{1999}).

\bibitem{Fukuma:2008_APL}
\bibinfo{author}{Fukuma, Y.} \emph{et~al.}
\newblock \bibinfo{title}{Carrier-induced ferromagnetism in
  {Ge$_{0.92}$Mn$_{0.08}$Te} epilayers with a {Curie} temperature up to 190
  {K}}.
\newblock \emph{\bibinfo{journal}{Appl. Phys. Lett.}}
  \textbf{\bibinfo{volume}{93}}, \bibinfo{pages}{252502}
  (\bibinfo{year}{2008}).

\bibitem{Olejnik:2008_PRB}
\bibinfo{author}{{Olejn{\'i}k}, K.} \emph{et~al.}
\newblock \bibinfo{title}{{Enhanced annealing, high {Curie} temperature and
  low-voltage gating in (Ga,Mn)As: A surface oxide control study}}.
\newblock \emph{\bibinfo{journal}{Phys. Rev. B}} \textbf{\bibinfo{volume}{78}},
  \bibinfo{pages}{054403} (\bibinfo{year}{2008}).

\bibitem{Wang:2008_APL}
\bibinfo{author}{Wang, M.} \emph{et~al.}
\newblock \bibinfo{title}{Achieving high {Curie} temperature in {(Ga,Mn)As}}.
\newblock \emph{\bibinfo{journal}{Appl. Phys. Lett.}}
  \textbf{\bibinfo{volume}{93}}, \bibinfo{pages}{132103}
  (\bibinfo{year}{2008}).

\bibitem{Chen:2009_APL}
\bibinfo{author}{Chen, L.} \emph{et~al.}
\newblock \bibinfo{title}{Low-temperature magnetotransport behaviors of heavily
  {Mn}-doped {(Ga,Mn)As} films with high ferromagnetic transition temperature}.
\newblock \emph{\bibinfo{journal}{Appl. Phys. Lett.}}
  \textbf{\bibinfo{volume}{95}}, \bibinfo{pages}{182505}
  (\bibinfo{year}{2009}).

\bibitem{Akai:1998_PRL}
\bibinfo{author}{Akai, H.}
\newblock \bibinfo{title}{Ferromagnetism and its stability in the diluted
  magnetic semiconductor {(In,Mn)As}}.
\newblock \emph{\bibinfo{journal}{Phys. Rev. Lett.}}
  \textbf{\bibinfo{volume}{81}}, \bibinfo{pages}{3002--3005}
  (\bibinfo{year}{1998}).

\bibitem{Sato:2003_EPL}
\bibinfo{author}{Sato, K.}, \bibinfo{author}{Dederics, P.~H.} \&
  \bibinfo{author}{Katayama-Yoshida, H.}
\newblock \bibinfo{title}{{Curie temperatures of {III-V} diluted magnetic
  semiconductors calculated from first principles}}.
\newblock \emph{\bibinfo{journal}{{Europhys. Lett.}}}
  \textbf{\bibinfo{volume}{{61}}}, \bibinfo{pages}{{403--408}}
  (\bibinfo{year}{2003}).

\bibitem{Mahadevan:2004_APL}
\bibinfo{author}{Mahadevan, P.} \& \bibinfo{author}{Zunger, A.}
\newblock \bibinfo{title}{Trends in ferromagnetism, hole localization, and
  acceptor level depth for {Mn} substitution in {GaN}, {GaP}, {GaAs}, {GaSb}}.
\newblock \emph{\bibinfo{journal}{Appl. Phys. Lett.}}
  \textbf{\bibinfo{volume}{85}}, \bibinfo{pages}{2860--2862}
  (\bibinfo{year}{2004}).

\bibitem{Burch:2008_JMMM}
\bibinfo{author}{Burch, K.}, \bibinfo{author}{Awschalom, D.} \&
  \bibinfo{author}{Basov, D.}
\newblock \bibinfo{title}{Optical properties of {III-Mn-V} ferromagnetic
  semiconductors}.
\newblock \emph{\bibinfo{journal}{J. Magn. Magn. Mat.}}
  \textbf{\bibinfo{volume}{320}}, \bibinfo{pages}{3207--3228}
  (\bibinfo{year}{2008}).

\bibitem{Alberi:2008_PRB}
\bibinfo{author}{Alberi, K.} \emph{et~al.}
\newblock \bibinfo{title}{Formation of {Mn}-derived impurity band in {III-Mn-V}
  alloys by valence band anticrossing}.
\newblock \emph{\bibinfo{journal}{Phys. Rev. B}} \textbf{\bibinfo{volume}{78}},
  \bibinfo{pages}{075201} (\bibinfo{year}{2008}).

\bibitem{Liu:2005_JMSME}
\bibinfo{author}{Liu, C.}, \bibinfo{author}{Yun, F.} \&
  \bibinfo{author}{Morko\c{c}, H.}
\newblock \bibinfo{title}{Ferromagnetism of {ZnO} and {GaN}: {A Review}}.
\newblock \emph{\bibinfo{journal}{J. Mater. Sci.:Mater. Electron.}}
  \textbf{\bibinfo{volume}{16}}, \bibinfo{pages}{555--597}
  (\bibinfo{year}{2005}).

\bibitem{Coey:2006_COSSMS}
\bibinfo{author}{Coey, J. M.~D.}
\newblock \bibinfo{title}{Dilute magnetic oxides}.
\newblock \emph{\bibinfo{journal}{Current Opinion Solid State Mater. Sci.}}
  \textbf{\bibinfo{volume}{10}}, \bibinfo{pages}{83--92}
  (\bibinfo{year}{2006}).

\bibitem{Bonanni:2007_SST}
\bibinfo{author}{Bonanni, A.}
\newblock \bibinfo{title}{Ferromagnetic nitride-based semiconductors doped with
  transition metals and rare earths}.
\newblock \emph{\bibinfo{journal}{Semicond. Sci. Technol.}}
  \textbf{\bibinfo{volume}{22}}, \bibinfo{pages}{R41--R56}
  (\bibinfo{year}{2007}).

\bibitem{Sato:2010_RMP}
\bibinfo{author}{Sato, K.} \emph{et~al.}
\newblock \bibinfo{title}{First-principles theory of dilute magnetic
  semiconductors}.
\newblock \emph{\bibinfo{journal}{Rev. Mod. Phys.}}
  \textbf{\bibinfo{volume}{82}}, \bibinfo{pages}{1633--1690}
  (\bibinfo{year}{2010}).

\bibitem{Blinowski:1996_PRB}
\bibinfo{author}{Blinowski, J.}, \bibinfo{author}{Kacman, P.} \&
  \bibinfo{author}{Majewski, J.~A.}
\newblock \bibinfo{title}{Ferromagnetic superexchange in {Cr}-based diluted
  magnetic semiconductors}.
\newblock \emph{\bibinfo{journal}{Phys. Rev. B}} \textbf{\bibinfo{volume}{53}},
  \bibinfo{pages}{9524--9527} (\bibinfo{year}{1996}).

\bibitem{Walsh:2008_PRL}
\bibinfo{author}{Walsh, A.}, \bibinfo{author}{{Da Silva}, J. L.~F.} \&
  \bibinfo{author}{Wei, S.-H.}
\newblock \bibinfo{title}{Theoretical description of carrier mediated magnetism
  in {Cobalt} doped {ZnO}}.
\newblock \emph{\bibinfo{journal}{Phys. Rev. Lett.}}
  \textbf{\bibinfo{volume}{100}}, \bibinfo{pages}{256401}
  (\bibinfo{year}{2008}).

\bibitem{Coey:2005_NM}
\bibinfo{author}{Coey, J. M.~D.}, \bibinfo{author}{Venkatesan, M.} \&
  \bibinfo{author}{Fitzgerald, C.~B.}
\newblock \bibinfo{title}{Donor impurity band exchange in dilute ferromagnetic
  oxides}.
\newblock \emph{\bibinfo{journal}{Nature Mater.}} \textbf{\bibinfo{volume}{4}},
  \bibinfo{pages}{173�179} (\bibinfo{year}{2005}).

\bibitem{Wang:2009_PRB}
\bibinfo{author}{Wang, Q.}, \bibinfo{author}{Sun, Q.}, \bibinfo{author}{Jena,
  P.} \& \bibinfo{author}{Kawazoe, Y.}
\newblock \bibinfo{title}{Magnetic properties of transition-metal-doped {Zn$_{1
  - x}$T$_x$O (T = Cr, Mn, Fe, Co, and Ni)} thin films with and without
  intrinsic defects: A density functional study}.
\newblock \emph{\bibinfo{journal}{Phys. Rev. B}} \textbf{\bibinfo{volume}{79}},
  \bibinfo{pages}{115407} (\bibinfo{year}{2009}).

\bibitem{Park:2005_PRL}
\bibinfo{author}{Park, C.~H.} \& \bibinfo{author}{Chadi, D.~J.}
\newblock \bibinfo{title}{Hydrogen-mediated spin-spin interaction in {ZnCoO}}.
\newblock \emph{\bibinfo{journal}{Phys. Rev. Lett.}}
  \textbf{\bibinfo{volume}{94}}, \bibinfo{pages}{127204}
  (\bibinfo{year}{2005}).

\bibitem{Coey:2008_JPD}
\bibinfo{author}{Coey, J. M.~D.}, \bibinfo{author}{Wongsaprom, K.},
  \bibinfo{author}{Alaria, J.} \& \bibinfo{author}{Venkatesan, M.}
\newblock \bibinfo{title}{Charge-transfer ferromagnetism in oxide
  nanoparticles}.
\newblock \emph{\bibinfo{journal}{J. Phys. D: Appl. Phys.}}
  \textbf{\bibinfo{volume}{41}}, \bibinfo{pages}{134012}
  (\bibinfo{year}{2008}).

\bibitem{Dietl:2003_NM}
\bibinfo{author}{Dietl, T.}
\newblock \bibinfo{title}{Dilute magnetic semiconductors: Functional
  ferromagnets}.
\newblock \emph{\bibinfo{journal}{Nature Mater.}} \textbf{\bibinfo{volume}{2}},
  \bibinfo{pages}{646--648} (\bibinfo{year}{2003}).

\bibitem{Cho:2008_APL}
\bibinfo{author}{Cho, Y.~J.}, \bibinfo{author}{Yu, K.~M.},
  \bibinfo{author}{Liu, X.}, \bibinfo{author}{Walukiewicz, W.} \&
  \bibinfo{author}{Furdyna, J.~K.}
\newblock \bibinfo{title}{Effects of donor doping on {Ga$_{1-x}$Mn$_x$As}}.
\newblock \emph{\bibinfo{journal}{Appl. Phys. Lett.}}
  \textbf{\bibinfo{volume}{93}}, \bibinfo{pages}{262505}
  (\bibinfo{year}{2008}).

\bibitem{Mayer:2010_PRB}
\bibinfo{author}{Mayer, M.~A.} \emph{et~al.}
\newblock \bibinfo{title}{Electronic structure of {Ga$_{1-x}$Mn$_x$As} analyzed
  according to hole-concentration-dependent measurements}.
\newblock \emph{\bibinfo{journal}{Phys. Rev. B}} \textbf{\bibinfo{volume}{81}},
  \bibinfo{pages}{045205} (\bibinfo{year}{2010}).

\bibitem{Jungwirth:2007_PRB}
\bibinfo{author}{Jungwirth, T.} \emph{et~al.}
\newblock \bibinfo{title}{Character of states near the {Fermi} level in
  {(Ga,Mn)As}: Impurity to valence band crossover}.
\newblock \emph{\bibinfo{journal}{Phys. Rev. B}} \textbf{\bibinfo{volume}{76}},
  \bibinfo{pages}{125206} (\bibinfo{year}{2007}).

\bibitem{Dietl:2008_JPSJ}
\bibinfo{author}{Dietl, T.}
\newblock \bibinfo{title}{{Interplay between carrier localization and magnetism
  in diluted magnetic and ferromagnetic semiconductors}}.
\newblock \emph{\bibinfo{journal}{J. Phys. Soc. Jpn.}}
  \textbf{\bibinfo{volume}{77}}, \bibinfo{pages}{031005}
  (\bibinfo{year}{2008}).

\bibitem{Altshuler:1985_B}
\bibinfo{author}{Altshuler, B.~L.} \& \bibinfo{author}{Aronov, A.~G.}
\newblock In \bibinfo{editor}{Efros, A.~L.} \& \bibinfo{editor}{Pollak, M.}
  (eds.) \emph{\bibinfo{booktitle}{Electron-Electron Interactions in Disordered
  Systems}}, \bibinfo{pages}{1} (\bibinfo{publisher}{North Holland, Amsterdam},
  \bibinfo{year}{1985}).

\bibitem{Lee:1985_RMP}
\bibinfo{author}{Lee, P.~A.} \& \bibinfo{author}{Ramakrishnan, T.~V.}
\newblock \bibinfo{title}{Disordered electronic systems}.
\newblock \emph{\bibinfo{journal}{Rev. Mod. Phys.}}
  \textbf{\bibinfo{volume}{57}}, \bibinfo{pages}{287--337}
  (\bibinfo{year}{1985}).

\bibitem{Belitz:1994_RMP}
\bibinfo{author}{Belitz, D.} \& \bibinfo{author}{Kirkpatrick, T.~R.}
\newblock \bibinfo{title}{{The Anderson-Mott transition}}.
\newblock \emph{\bibinfo{journal}{Rev. Mod. Phys.}}
  \textbf{\bibinfo{volume}{66}}, \bibinfo{pages}{261--380}
  (\bibinfo{year}{1994}).

\bibitem{Edwards:1978_PRB}
\bibinfo{author}{Edwards, P.~P.} \& \bibinfo{author}{Sienko, M.~J.}
\newblock \bibinfo{title}{Universality aspects of the metal-nonmetal transition
  in condensed media}.
\newblock \emph{\bibinfo{journal}{Phys. Rev. B}} \textbf{\bibinfo{volume}{17}},
  \bibinfo{pages}{2575--2581} (\bibinfo{year}{1978}).

\bibitem{Dietl:2002_PRB}
\bibinfo{author}{Dietl, T.}, \bibinfo{author}{Matsukura, F.} \&
  \bibinfo{author}{Ohno, H.}
\newblock \bibinfo{title}{{Ferromagnetism of magnetic semiconductors:
  Zhang-Rice limit}}.
\newblock \emph{\bibinfo{journal}{Phys. Rev. B}} \textbf{\bibinfo{volume}{66}},
  \bibinfo{pages}{033203} (\bibinfo{year}{2002}).

\bibitem{Dietl:2008_PRB}
\bibinfo{author}{Dietl, T.}
\newblock \bibinfo{title}{{Hole states in wide band-gap diluted magnetic
  semiconductors and oxides}}.
\newblock \emph{\bibinfo{journal}{Phys. Rev. B}} \textbf{\bibinfo{volume}{77}},
  \bibinfo{pages}{085208} (\bibinfo{year}{2008}).

\bibitem{Neumaier:2009_PRL}
\bibinfo{author}{Neumaier, D.} \emph{et~al.}
\newblock \bibinfo{title}{All-electrical measurement of the density of states
  in {(Ga,Mn)As}}.
\newblock \emph{\bibinfo{journal}{Phys. Rev. Lett.}}
  \textbf{\bibinfo{volume}{103}}, \bibinfo{pages}{087203}
  (\bibinfo{year}{2009}).

\bibitem{Boukari:2002_PRL}
\bibinfo{author}{Boukari, H.} \emph{et~al.}
\newblock \bibinfo{title}{{Light and Electric Field Control of Ferromagnetism
  in Magnetic Quantum Structures}}.
\newblock \emph{\bibinfo{journal}{Phys. Rev. Lett.}}
  \textbf{\bibinfo{volume}{88}}, \bibinfo{pages}{207204}
  (\bibinfo{year}{2002}).

\bibitem{Nishitani:2010_PRB}
\bibinfo{author}{Nishitani, Y.} \emph{et~al.}
\newblock \bibinfo{title}{Curie temperature versus hole concentration in
  field-effect structures of {Ga$_{1-x}$Mn$_x$As}}.
\newblock \emph{\bibinfo{journal}{Phys. Rev. B}} \textbf{\bibinfo{volume}{81}},
  \bibinfo{pages}{045208} (\bibinfo{year}{2010}).

\bibitem{MacDonald:2005_NM}
\bibinfo{author}{MacDonald, A.~H.}, \bibinfo{author}{Schiffer, P.} \&
  \bibinfo{author}{Samarth, N.}
\newblock \bibinfo{title}{Ferromagnetic semiconductors: moving beyond
  {(Ga,Mn)As}}.
\newblock \emph{\bibinfo{journal}{Nature Mater.}} \textbf{\bibinfo{volume}{4}},
  \bibinfo{pages}{195--202} (\bibinfo{year}{2005}).

\bibitem{Edmonds:2002_APL}
\bibinfo{author}{Edmonds, K.~W.} \emph{et~al.}
\newblock \bibinfo{title}{{High} {Curie} temperature {GaMnAs} obtained by
  resistance-monitored annealing}.
\newblock \emph{\bibinfo{journal}{Appl. Phys. Lett.}}
  \textbf{\bibinfo{volume}{81}}, \bibinfo{pages}{4991--4993}
  (\bibinfo{year}{2002}).

\bibitem{Cho:2008_SST}
\bibinfo{author}{Cho, Y.~J.}, \bibinfo{author}{Liu, X.} \&
  \bibinfo{author}{Furdyna, J.~K.}
\newblock \bibinfo{title}{Collapse of ferromagnetism in {(Ga, Mn)As} at high
  hole concentrations}.
\newblock \emph{\bibinfo{journal}{Semicond. Sci. Technol.}}
  \textbf{\bibinfo{volume}{23}}, \bibinfo{pages}{125010}
  (\bibinfo{year}{2008}).

\bibitem{Furdyna:2004_JPCM}
\bibinfo{author}{Furdyna, J.~K.} \emph{et~al.}
\newblock \bibinfo{title}{Fermi level effects on {Mn} incorporation in
  modulation-doped ferromagnetic {III$_{1-x}$Mn$_x$V} heterostructures}.
\newblock \emph{\bibinfo{journal}{J. Phys.: Condens. Matter}}
  \textbf{\bibinfo{volume}{16}}, \bibinfo{pages}{S5499–--S5508}
  (\bibinfo{year}{2004}).

\bibitem{Scarpulla:2005_PRL}
\bibinfo{author}{Scarpulla, M.~A.} \emph{et~al.}
\newblock \bibinfo{title}{{Ferromagnetism in {Ga$_{1-x}$Mn$_{x}$P:} evidence
  for inter-Mn exchange mediated by localized holes within a detached impurity
  band}}.
\newblock \emph{\bibinfo{journal}{Phys. Rev. Lett.}}
  \textbf{\bibinfo{volume}{95}}, \bibinfo{pages}{207204}
  (\bibinfo{year}{2005}).

\bibitem{Schallenberg:2006_APL}
\bibinfo{author}{Schallenberg, T.} \& \bibinfo{author}{Munekata, H.}
\newblock \bibinfo{title}{Preparation of ferromagnetic {(In,Mn)As} with a high
  {Curie} temperature of 90 {K}}.
\newblock \emph{\bibinfo{journal}{Appl. Phys. Lett.}}
  \textbf{\bibinfo{volume}{89}}, \bibinfo{pages}{042507}
  (\bibinfo{year}{2006}).

\bibitem{Abe:2000_PE}
\bibinfo{author}{Abe, E.}, \bibinfo{author}{Matsukura, F.},
  \bibinfo{author}{Yasuda, H.}, \bibinfo{author}{Ohno, Y.} \&
  \bibinfo{author}{Ohno, H.}
\newblock \bibinfo{title}{{Molecular beam epitaxy of III-V diluted magnetic
  semiconductor (Ga,Mn)Sb}}.
\newblock \emph{\bibinfo{journal}{Physica}} \textbf{\bibinfo{volume}{E 7}},
  \bibinfo{pages}{981--985} (\bibinfo{year}{2000}).

\bibitem{Wojtowicz:2003_APL}
\bibinfo{author}{Wojtowicz, T.} \emph{et~al.}
\newblock \bibinfo{title}{{In$_{1-x}$Mn$_{x}$Sb --- a new narrow gap
  ferromagnetic semiconductor}}.
\newblock \emph{\bibinfo{journal}{Appl. Phys. Lett.}}
  \textbf{\bibinfo{volume}{82}}, \bibinfo{pages}{4310--4312}
  (\bibinfo{year}{2003}).

\bibitem{Jungwirth:2002_PRB}
\bibinfo{author}{Jungwirth, T.}, \bibinfo{author}{{K{\"o}nig}, J.},
  \bibinfo{author}{Sinova, J.}, \bibinfo{author}{{Ku\v{c}era}, J.} \&
  \bibinfo{author}{MacDonald, A.~H.}
\newblock \bibinfo{title}{{Curie} temperature trends in {(III,Mn)V}
  ferromagnetic semiconductors}.
\newblock \emph{\bibinfo{journal}{Phys. Rev. B}} \textbf{\bibinfo{volume}{66}},
  \bibinfo{pages}{012402} (\bibinfo{year}{2002}).

\bibitem{Jungwirth:2006_RMP}
\bibinfo{author}{Jungwirth, T.}, \bibinfo{author}{Sinova, J.},
  \bibinfo{author}{{Ma\v{s}ek}, J.}, \bibinfo{author}{{Ku\v{c}era}, J.} \&
  \bibinfo{author}{MacDonald, A.~H.}
\newblock \bibinfo{title}{Theory of ferromagnetic {(III,Mn)V} semiconductors}.
\newblock \emph{\bibinfo{journal}{Rev. Mod. Phys.}}
  \textbf{\bibinfo{volume}{78}}, \bibinfo{pages}{809--864}
  (\bibinfo{year}{2006}).

\bibitem{Glunk:2009_PRB}
\bibinfo{author}{Glunk, M.} \emph{et~al.}
\newblock \bibinfo{title}{Magnetic anisotropy in {(Ga,Mn)As}: Influence of
  epitaxial strain and hole concentration}.
\newblock \emph{\bibinfo{journal}{Phys. Rev. B}} \textbf{\bibinfo{volume}{79}},
  \bibinfo{pages}{195206} (\bibinfo{year}{2009}).

\bibitem{Kodzuka:2009_U}
\bibinfo{author}{Kodzuka, M.}, \bibinfo{author}{Ohkubo, T.},
  \bibinfo{author}{Hono, K.}, \bibinfo{author}{Matsukura, F.} \&
  \bibinfo{author}{Ohno, H.}
\newblock \bibinfo{title}{{3DAP analysis of {(Ga,Mn)As} diluted magnetic
  semiconductor thin film}}.
\newblock \emph{\bibinfo{journal}{Ultramicroscopy}}
  \textbf{\bibinfo{volume}{109}}, \bibinfo{pages}{644--648}
  (\bibinfo{year}{2009}).

\bibitem{Yu:2002_PRB}
\bibinfo{author}{Yu, K.~M.} \emph{et~al.}
\newblock \bibinfo{title}{Effect of the location of {Mn} sites in ferromagnetic
  {Ga$_{1-x}$Mn$_{x}$As} on its {Curie} temperature}.
\newblock \emph{\bibinfo{journal}{Phys. Rev. B}} \textbf{\bibinfo{volume}{65}},
  \bibinfo{pages}{201303} (\bibinfo{year}{2002}).

\bibitem{Wierzbowska:2004_PRB}
\bibinfo{author}{Wierzbowska, M.}, \bibinfo{author}{Sanchez-Portal, D.} \&
  \bibinfo{author}{Sanvito, S.}
\newblock \bibinfo{title}{Different origin of the ferromagnetic order in
  {(Ga,Mn)As} and {(Ga,Mn)N}}.
\newblock \emph{\bibinfo{journal}{Phys. Rev. B}} \textbf{\bibinfo{volume}{70}},
  \bibinfo{pages}{235209} (\bibinfo{year}{2004}).

\bibitem{Schulthess:2005_NM}
\bibinfo{author}{Schulthess, T.}, \bibinfo{author}{Temmerman, W.~M.},
  \bibinfo{author}{Szotek, Z.}, \bibinfo{author}{Butler, W.~H.} \&
  \bibinfo{author}{Stocks, G.~M.}
\newblock \bibinfo{title}{Electronic {Structure} and {Exchange} {Coupling} of
  {Mn} {Impurities} in {III-V} {Semiconductors}}.
\newblock \emph{\bibinfo{journal}{Nature Mater.}} \textbf{\bibinfo{volume}{4}},
  \bibinfo{pages}{838--844} (\bibinfo{year}{2005}).

\bibitem{Kitchen:2006_N}
\bibinfo{author}{Kitchen, D.}, \bibinfo{author}{Richardella, A.},
  \bibinfo{author}{Tang, J.-M.}, \bibinfo{author}{Flatte, M.~E.} \&
  \bibinfo{author}{Yazdani, A.}
\newblock \bibinfo{title}{Atom-by-atom substitution of {Mn} in {GaAs} and
  visualization of their hole-mediated interactions}.
\newblock \emph{\bibinfo{journal}{Nature}} \textbf{\bibinfo{volume}{442}},
  \bibinfo{pages}{436} (\bibinfo{year}{2006}).

\bibitem{Sarigiannidou:2006_PRB}
\bibinfo{author}{Sarigiannidou, E.} \emph{et~al.}
\newblock \bibinfo{title}{Intrinsic ferromagnetism in wurtzite {(Ga,Mn)N}
  semiconductor}.
\newblock \emph{\bibinfo{journal}{Phys. Rev. B}} \textbf{\bibinfo{volume}{74}},
  \bibinfo{pages}{041306} (\bibinfo{year}{2006}).

\bibitem{Martinez-Criado:2005_APL}
\bibinfo{author}{Martinez-Criado, G.} \emph{et~al.}
\newblock \bibinfo{title}{{Mn-rich clusters in {GaN}: {Hexagonal} or cubic
  symmetry?}}
\newblock \emph{\bibinfo{journal}{Appl. Phys. Lett.}}
  \textbf{\bibinfo{volume}{86}}, \bibinfo{pages}{131927}
  (\bibinfo{year}{2005}).

\bibitem{Lipinska:2009_PRB}
\bibinfo{author}{Lipi\'{n}ska, A.} \emph{et~al.}
\newblock \bibinfo{title}{Ferromagnetic properties of p-{(Cd,Mn)Te} quantum
  wells: {Interpretation} of magneto-optical measurements by {Monte} {Carlo}
  simulations}.
\newblock \emph{\bibinfo{journal}{Phys. Rev. B}} \textbf{\bibinfo{volume}{79}},
  \bibinfo{pages}{235322} (\bibinfo{year}{2009}).

\bibitem{Ferrand:2001_PRB}
\bibinfo{author}{Ferrand, D.} \emph{et~al.}
\newblock \bibinfo{title}{Carrier-induced ferromagnetism in
  {p-Zn$_{1-x}$Mn$_{x}$Te}}.
\newblock \emph{\bibinfo{journal}{Phys. Rev. B}} \textbf{\bibinfo{volume}{63}},
  \bibinfo{pages}{085201} (\bibinfo{year}{2001}).

\bibitem{Andrearczyk:2001_ICPS}
\bibinfo{author}{Andrearczyk, T.} \emph{et~al.}
\newblock \bibinfo{title}{Ferromagnetic interactions in p- and n-type {II-VI}
  diluted magnetic semiconductors}.
\newblock In \bibinfo{editor}{Miura, N.} \& \bibinfo{editor}{Ando, T.} (eds.)
  \emph{\bibinfo{booktitle}{Proceedings 25th International Conference on
  Physics of Semiconductors, Osaka, Japan, 2000}}, \bibinfo{pages}{235--236}
  (\bibinfo{publisher}{Springer, Berlin}, \bibinfo{year}{2001}).

\bibitem{Serrate:2007_JPCM}
\bibinfo{author}{Serrate, D.}, \bibinfo{author}{Teresa, J.~M.~D.} \&
  \bibinfo{author}{Ibarra, M.~R.}
\newblock \bibinfo{title}{{Double perovskites with ferromagnetism above room
  temperature}}.
\newblock \emph{\bibinfo{journal}{J. Phys.: Cond. Matter}}
  \textbf{\bibinfo{volume}{19}}, \bibinfo{pages}{023201}
  (\bibinfo{year}{2007}).

\bibitem{Park:2004_APL}
\bibinfo{author}{Park, J.~H.}, \bibinfo{author}{Kim, M.~G.},
  \bibinfo{author}{Jang, H.~M.}, \bibinfo{author}{Ryu, S.} \&
  \bibinfo{author}{Kim, Y.~M.}
\newblock \bibinfo{title}{Co-metal clustering as the origin of ferromagnetism
  in {Co}-doped {ZnO} thin films}.
\newblock \emph{\bibinfo{journal}{Appl. Phys. Lett.}}
  \textbf{\bibinfo{volume}{84}}, \bibinfo{pages}{1338--1340}
  (\bibinfo{year}{2004}).

\bibitem{Bonanni:2007_PRB}
\bibinfo{author}{Bonanni, A.} \emph{et~al.}
\newblock \bibinfo{title}{Paramagnetic {GaN:Fe} and ferromagnetic {(Ga,Fe)N}:
  {The} relationship between structural, electronic, and magnetic properties}.
\newblock \emph{\bibinfo{journal}{Phys. Rev. B}} \textbf{\bibinfo{volume}{75}},
  \bibinfo{pages}{125210} (\bibinfo{year}{2007}).

\bibitem{De_Boeck:1996_APL}
\bibinfo{author}{Boeck, J.~D.} \emph{et~al.}
\newblock \bibinfo{title}{Nanometer-scale magnetic {MnAs} particles in {GaAs}
  grown by molecular beam epitaxy}.
\newblock \emph{\bibinfo{journal}{App. Phys. Lett.}}
  \textbf{\bibinfo{volume}{68}}, \bibinfo{pages}{2744--2746}
  (\bibinfo{year}{1996}).

\bibitem{Moreno:2002_JAP}
\bibinfo{author}{Moreno, M.}, \bibinfo{author}{Trampert, A.},
  \bibinfo{author}{Jenichen, B.}, \bibinfo{author}{{D{\"a}weritz}, L.} \&
  \bibinfo{author}{Ploog, K.~H.}
\newblock \bibinfo{title}{{Correlation of structure and magnetism in {GaAs}
  with embedded {Mn(Ga)As} magnetic nanoclusters}}.
\newblock \emph{\bibinfo{journal}{J. Appl. Phys.}}
  \textbf{\bibinfo{volume}{92}}, \bibinfo{pages}{4672--4677}
  (\bibinfo{year}{2002}).

\bibitem{Rovezzi:2009_PRB}
\bibinfo{author}{Rovezzi, M.} \emph{et~al.}
\newblock \bibinfo{title}{{Local structure of (Ga,Fe)N and (Ga,Fe)N:Si
  investigated by x-ray absorption fine structure spectroscopy}}.
\newblock \emph{\bibinfo{journal}{Phys. Rev. B}} \textbf{\bibinfo{volume}{79}},
  \bibinfo{pages}{195209} (\bibinfo{year}{2009}).

\bibitem{Bonanni:2008_PRL}
\bibinfo{author}{Bonanni, A.} \emph{et~al.}
\newblock \bibinfo{title}{{Controlled Aggregation of Magnetic Ions in a
  Semiconductor: An Experimental Demonstration}}.
\newblock \emph{\bibinfo{journal}{Phys. Rev. Lett.}}
  \textbf{\bibinfo{volume}{101}}, \bibinfo{pages}{135502}
  (\bibinfo{year}{2008}).

\bibitem{Tanaka:2008_B}
\bibinfo{author}{Tanaka, M.}, \bibinfo{author}{Yokoyama, M.},
  \bibinfo{author}{Hai, P.~N.} \& \bibinfo{author}{Ohya, S.}
\newblock \bibinfo{title}{{Properties and functionalities of {MnAs/III-V}
  hybrid and composite structures}}.
\newblock In \bibinfo{editor}{Dietl, T.}, \bibinfo{editor}{Awschalom, D.~D.},
  \bibinfo{editor}{Kaminska, M.} \& \bibinfo{editor}{Ohno, H.} (eds.)
  \emph{\bibinfo{booktitle}{Spintronics}}, \bibinfo{pages}{pp 455--485}
  (\bibinfo{publisher}{Elsevier, Amsterdam}, \bibinfo{year}{2008}).

\bibitem{Kuroda:2007_NM}
\bibinfo{author}{Kuroda, S.} \emph{et~al.}
\newblock \bibinfo{title}{Origin and control of high temperature ferromagnetism
  in semiconductors}.
\newblock \emph{\bibinfo{journal}{Nature Mater.}} \textbf{\bibinfo{volume}{6}},
  \bibinfo{pages}{440--446} (\bibinfo{year}{2007}).

\bibitem{Gu:2005_JMMM}
\bibinfo{author}{Gu, L.} \emph{et~al.}
\newblock \bibinfo{title}{{Characterization of Al(Cr)N and Ga(Cr)N dilute
  magnetic semiconductors}}.
\newblock \emph{\bibinfo{journal}{J. Magn. Magn. Mater.}}
  \textbf{\bibinfo{volume}{290-291}}, \bibinfo{pages}{1395--1397}
  (\bibinfo{year}{2005}).

\bibitem{Jamet:2006_NM}
\bibinfo{author}{Jamet, M.} \emph{et~al.}
\newblock \bibinfo{title}{{High-Curie-temperature ferromagnetism in
  self-organized {Ge$_{1-x}$Mn$_x$} nanocolumns}}.
\newblock \emph{\bibinfo{journal}{Nature Mater.}} \textbf{\bibinfo{volume}{5}},
  \bibinfo{pages}{653--659} (\bibinfo{year}{2006}).

\bibitem{Nishio:2009_MRSP}
\bibinfo{author}{Nishio, Y.}, \bibinfo{author}{Ishikawa, K.},
  \bibinfo{author}{Kuroda, S.}, \bibinfo{author}{Mitome, M.} \&
  \bibinfo{author}{Bando, Y.}
\newblock \bibinfo{title}{Formation of {Cr}-rich nano-clusters and columns in
  {(Zn,Cr)Te} grown by {MBE}}.
\newblock In \emph{\bibinfo{booktitle}{Mater. Res. Soc. Symp. Proc.}}, vol.
  \bibinfo{volume}{1183-FF01-11} of \emph{\bibinfo{series}{MRS Proceedings}}
  (\bibinfo{publisher}{Materials Research Society}, \bibinfo{year}{2009}).

\bibitem{Katayama-Yoshida:2007_PSSA}
\bibinfo{author}{Katayama-Yoshida, H.} \emph{et~al.}
\newblock \bibinfo{title}{Theory of ferromagnetic semiconductors}.
\newblock \emph{\bibinfo{journal}{phys.stat. sol. (a)}}
  \textbf{\bibinfo{volume}{204}}, \bibinfo{pages}{15--32}
  (\bibinfo{year}{2007}).

\bibitem{Dietl:2006_NM}
\bibinfo{author}{Dietl, T.}
\newblock \bibinfo{title}{{Self-Organised Growth Controlled by Charge States of
  Magnetic Impurities}}.
\newblock \emph{\bibinfo{journal}{Nature Mater.}} \textbf{\bibinfo{volume}{5}},
  \bibinfo{pages}{673} (\bibinfo{year}{2006}).

\bibitem{Ye:2006_PRB}
\bibinfo{author}{Ye, L.-H.} \& \bibinfo{author}{Freeman, A.~J.}
\newblock \bibinfo{title}{{Defect compensation, clustering, and magnetism in
  {Cr-doped} anatase}}.
\newblock \emph{\bibinfo{journal}{Phys. Rev. B}} \textbf{\bibinfo{volume}{73}},
  \bibinfo{pages}{081304(R)} (\bibinfo{year}{2006}).

\bibitem{Straumal:2009_PRB}
\bibinfo{author}{Straumal, B.~B.} \emph{et~al.}
\newblock \bibinfo{title}{Magnetization study of nanograined pure and
  {Mn}-doped {ZnO} films: Formation of a ferromagnetic grain-boundary foam}.
\newblock \emph{\bibinfo{journal}{Phys. Rev. B}} \textbf{\bibinfo{volume}{79}},
  \bibinfo{pages}{205206} (\bibinfo{year}{2009}).

\bibitem{Ney:2008_PRL}
\bibinfo{author}{Ney, A.} \emph{et~al.}
\newblock \bibinfo{title}{Absence of intrinsic ferromagnetic interactions of
  isolated and paired {Co} dopant atoms in {Zn$_{1-x}$Co$_x$O} with high
  structural perfection}.
\newblock \emph{\bibinfo{journal}{Phys. Rev. Lett.}}
  \textbf{\bibinfo{volume}{100}}, \bibinfo{pages}{157201}
  (\bibinfo{year}{2008}).

\bibitem{Schilfgaarde:2001_PRB}
\bibinfo{author}{van Schilfgaarde, M.} \& \bibinfo{author}{Mryasov, O.~N.}
\newblock \bibinfo{title}{{Anomalous exchange interactions in {III-V} dilute
  magnetic semiconductors}}.
\newblock \emph{\bibinfo{journal}{Phys. Rev. B}} \textbf{\bibinfo{volume}{63}},
  \bibinfo{pages}{233205} (\bibinfo{year}{2001}).

\bibitem{Zunger:1986_SSP}
\bibinfo{author}{Zunger, A.}
\newblock \bibinfo{title}{Electronic {Structure} of 3d {Transition-Atom}
  {Impurities} in {Semiconductors}}.
\newblock In \bibinfo{editor}{Seitz, F.} \& \bibinfo{editor}{Turnbull, D.}
  (eds.) \emph{\bibinfo{booktitle}{Solid State Physics}},
  vol.~\bibinfo{volume}{39}, \bibinfo{pages}{275--464}
  (\bibinfo{publisher}{Academic Press, New York}, \bibinfo{year}{1986}).

\bibitem{Furdyna:1988_B}
\bibinfo{editor}{Furdyna, J.~K.} \& \bibinfo{editor}{Kossut, J.} (eds.)
  \emph{\bibinfo{title}{{Diluted Magnetic Semiconductors}}},
  vol.~\bibinfo{volume}{25} of \emph{\bibinfo{series}{Semiconductors and
  Semimetals}} (\bibinfo{publisher}{Academic Press}, \bibinfo{address}{New
  York}, \bibinfo{year}{1988}).

\bibitem{Da_Silva:2008_NJP}
\bibinfo{author}{Da~Silva, J. L.~F.}, \bibinfo{author}{Dalpian, G.~M.} \&
  \bibinfo{author}{Wei, S.-H.}
\newblock \bibinfo{title}{{Carrier-induced enhancement and suppression of
  ferromagnetism in Zn$_{1-x}$Cr$_{x}$Te and Ga$_{1-x}$Cr$_{x}$As: origin of
  the spinodal decomposition}}.
\newblock \emph{\bibinfo{journal}{{New J. Phys.}}}
  \textbf{\bibinfo{volume}{{10}}}, \bibinfo{pages}{{113007}}
  (\bibinfo{year}{2008}).

\bibitem{Hai:2009_N}
\bibinfo{author}{Hai, P.~N.}, \bibinfo{author}{Ohya, S.},
  \bibinfo{author}{Tanaka, M.}, \bibinfo{author}{Barnes, S.~E.} \&
  \bibinfo{author}{Maekawa, S.}
\newblock \bibinfo{title}{Electromotive force and huge magnetoresistance in
  magnetic tunnel junctions}.
\newblock \emph{\bibinfo{journal}{Nature}} \textbf{\bibinfo{volume}{458}},
  \bibinfo{pages}{489--492} (\bibinfo{year}{2007}).

\bibitem{Katayama-Yoshida:2008_B}
\bibinfo{author}{Katayama-Yoshida, H.}, \bibinfo{author}{Sato, K.},
  \bibinfo{author}{Fukushima, T.}, \bibinfo{author}{M.~Toyoda, H.~K.} \&
  \bibinfo{author}{{An van Dinh}}.
\newblock \bibinfo{title}{Computational nano-materials design for the wide
  band-gap and high-{Tc} semiconductor spintronics}.
\newblock In \bibinfo{editor}{Dietl, T.}, \bibinfo{editor}{Awschalom, D.~D.},
  \bibinfo{editor}{Kaminska, M.} \& \bibinfo{editor}{Ohno, H.} (eds.)
  \emph{\bibinfo{booktitle}{Spintronics}}, \bibinfo{pages}{pp 433--454}
  (\bibinfo{publisher}{Elsevier, Amsterdam}, \bibinfo{year}{2008}).

\bibitem{Geshi:2005_AIPCP}
\bibinfo{author}{Geshi, M.}, \bibinfo{author}{Kusakabe, K.},
  \bibinfo{author}{Tsukamoto, H.} \& \bibinfo{author}{Suzuki, N.}
\newblock \bibinfo{title}{A new ferromagnetic material excluding transition
  metals: {CaAs} in a distorted zinc-blende structure}.
\newblock \emph{\bibinfo{journal}{AIP Conf. Proc.}}
  \textbf{\bibinfo{volume}{772}}, \bibinfo{pages}{327--328}
  (\bibinfo{year}{2005}).

\bibitem{Volnianska:2010_JPCM}
\bibinfo{author}{Volnianska, O.} \& \bibinfo{author}{Boguslawski, P.}
\newblock \bibinfo{title}{Magnetism of solids resulting from spin polarization
  of p orbitals}.
\newblock \emph{\bibinfo{journal}{J. Phys.: Condens. Matter}}
  \textbf{\bibinfo{volume}{22}}, \bibinfo{pages}{073202}
  (\bibinfo{year}{2010}).

\bibitem{Elfimov:2002_PRL}
\bibinfo{author}{Elfimov, I.~S.}, \bibinfo{author}{Yunoki, S.} \&
  \bibinfo{author}{Sawatzky, G.~A.}
\newblock \bibinfo{title}{Possible path to a new class of ferromagnetic and
  half-metallic ferromagnetic materials}.
\newblock \emph{\bibinfo{journal}{Phys. Rev. Lett.}}
  \textbf{\bibinfo{volume}{89}}, \bibinfo{pages}{216403}
  (\bibinfo{year}{2002}).

\bibitem{Sliwa:2008_PRB}
\bibinfo{author}{{\'S}liwa, C.} \& \bibinfo{author}{Dietl, T.}
\newblock \bibinfo{title}{Electron-hole contribution to the apparent $s - d$
  exchange interaction in {III-V} dilute magnetic semiconductors}.
\newblock \emph{\bibinfo{journal}{Phys. Rev. B}} \textbf{\bibinfo{volume}{78}},
  \bibinfo{pages}{165205} (\bibinfo{year}{2008}).

\bibitem{Grace:2009_AM}
\bibinfo{author}{Grace, P.~J.} \emph{et~al.}
\newblock \bibinfo{title}{The origin of the magnetism of etched {Silicon}}.
\newblock \emph{\bibinfo{journal}{{Adv. Mater.}}}
  \textbf{\bibinfo{volume}{{21}}}, \bibinfo{pages}{{71--74}}
  (\bibinfo{year}{2009}).

\end{thebibliography}


\end{document}